\documentclass[11pt, letterpaper]{article}
\usepackage{textcomp}
\usepackage{jheppub}
\usepackage{epsfig}
\usepackage{amsfonts}
\usepackage{amsmath}
\usepackage{amssymb}
\usepackage{dsfont}
\usepackage{hyperref}
\usepackage[dvipsnames,svgnames]{xcolor}
\usepackage{bbm}
\usepackage[most]{tcolorbox}
\usepackage{slashed}
\usepackage{booktabs, dcolumn, multicol, multirow, makecell}

\newcounter{cutbox}
    \renewcommand{\thecutbox}{\Roman{cutbox}}

\newcommand{\ep}{\epsilon}

\newcommand{\dslash}[1]{#1 \llap{/\kern-0.5pt}}
\newcommand{\Dslash}[1]{#1 \llap{/\kern+1.5pt}}
\newcommand{\DDslash}[1]{#1 \llap{/\kern+2.3pt}}
\newcommand{\dslashh}[1]{#1 \llap{/\kern+1pt}}

\newcommand{\beq}{\begin{equation}}
\newcommand{\eeq}{\end{equation}}

\newcommand{\bea}{\begin{eqnarray}}
\newcommand{\eea}{\end{eqnarray}}
\newcommand{\bma}{\begin{pmatrix}}
\newcommand{\ema}{\end{pmatrix}}
\newcommand{\nn}{\nonumber}

\title{Flavorful Lepton Number Violation at the EIC}

\author[a,b]{Sebastián Urrutia Quiroga,}
\author[a]{Vincenzo Cirigliano,}
\author[a]{Wouter Dekens,}
\author[c]{Kaori Fuyuto,}
\author[d]{Emanuele Mereghetti}

\affiliation[a]{Institute for Nuclear Theory, University of Washington, Seattle, WA 91195-1550, USA}
\affiliation[b]{Northwestern University, Department of Physics \& Astronomy, 2145 Sheridan Road, Evanston, IL 60208, USA}
\affiliation[c]{KEK Theory Center, IPNS, KEK, Tsukuba 305–0801, Japan}
\affiliation[d]{Theoretical Division, Los Alamos National Laboratory,
Los Alamos, NM 87545, USA}

\emailAdd{suq90@uw.edu}
\emailAdd{cirigv@uw.edu}
\emailAdd{wdekens@uw.edu}
\emailAdd{kfuyuto@post.kek.jp}
\emailAdd{emereghetti@lanl.gov}

\abstract{We explore the prospects of detecting flavorful lepton number violation at the Electron-Ion Collider (EIC) 
through resonant production of  heavy neutral leptons (HNLs), 
resulting in   $e^- p \to  \ell^+_\alpha + k\, j+X$, where  $\alpha \in \{e, \mu, \tau \}$ and $k$ 
denotes the number of jets.   
We work in the $\nu$SMEFT framework of the Standard Model Effective Field Theory 
augmented with $n$ singlet HNLs,  one of which is in the mass range $10-100$~GeV,  within kinematic reach of the EIC.   
To explore the EIC sensitivity, we focus on the HNL production mechanism induced by mixing with light neutrinos. 
We study kinematic distributions for signal and backgrounds,  including hadronization and detector effects, 
and suggest a set of cuts to minimize backgrounds.  
In the mass range considered, we find that the EIC with muon detection capabilities and  an  integrated luminosity of  
$100~\mathrm{fb}^{-1}$  can reach sensitivities comparable to the strongest direct (LHC) and indirect constraints,  
and is especially 
relevant in the $\nu$SMEFT framework beyond dimension four.
Our  study  motivates further  assessment of    muon detection capabilities at the EIC  and $\tau$ hadronic reconstruction, 
as well as  a more general theoretical analysis involving production mechanisms mediated by 
higher-dimensional operators in the effective theory.}

\begin{document}
{\flushright INT-PUB-26-006, KEK-TH-2808
\\[-9ex]
}

\maketitle

\section{Introduction}

Probes of  lepton number violation (LNV)  allow one to  explore the possible Majorana origin  of neutrino mass and, more generally, physics beyond the 
Standard Model (BSM)~\cite{Weinberg:1979sa,FileviezPerez:2022ypk}.   While neutrinoless double beta  ($0 \nu \beta \beta$) decay often provides the highest sensitivity to $\Delta L=2$ interactions~\cite{Agostini:2022zub}, in certain new-physics scenarios other LNV processes, such as lepton and meson decays 
or high-energy $ep$ or $pp$  collisions, can offer competitive sensitivity~\cite{FileviezPerez:2022ypk}.  
In broad brush, a given process can compete with $0\nu \beta \beta$ decay or can provide complementary information  when 
(i) the LNV process involves particles that can be produced  nearly on-shell, thereby overcoming with resonant enhancements the large effective fluxes corresponding to 100's of kilograms of decaying material in $0\nu \beta \beta$ decay;
(ii) the process involves LNV beyond the electron flavor sector. Examples of this second possibility include  $\mu^- \to e^+$ conversion in nuclei, 
$\tau^- \to \mu^+ \pi^- \pi^-$, $pp \to \ell^\pm_\alpha \ell^\pm_\beta + X$,  $e^- p \to \ell^+_\alpha +X$ , with $\alpha, \beta \in \{\mu, \tau \}$.

A large class of extensions of the Standard Model (SM) that generate neutrino masses involves new spin-1/2  Majorana fields that are singlets under the SM gauge group~\cite{Minkowski:1977sc, Gell-Mann:1979vob, Yanagida:1979as, Glashow:1979nm, Mohapatra:1979ia},   called `right-handed neutrinos', `sterile neutrinos',  or `heavy neutral leptons' (HNLs) in the literature. 
Depending on the mass scale associated with the new singlet fields,  models with HNLs can be probed in a broad array of processes that conserve or violate lepton number (see  Ref.~\cite{Abdullahi:2022jlv} for an overview).  

In this study, we will focus on the role that the Electron-Ion-Collider (EIC)~\cite{AbdulKhalek:2021gbh}
can play in probing  LNV originating from HNLs 
in the mass range 10-100~GeV. 
In this mass range, the leading contribution to LNV signals
 arises from resonant production of the HNL, $N$, in $e^- p \to N\,j + X$ followed by $N \to \ell^+_\alpha + W^*$    ($\alpha \in \{e,\mu, \tau \}$) 
and $W^*$ going into hadrons. 
 This channel has previously been studied in the context of HERA~\cite{Buchmuller:1991tu,Buchmuller:1992wm,Ingelman:1993ve},  
 the LHeC~\cite{Blaksley:2011ey,Duarte:2014zea,Mondal:2015zba,Lindner:2016lxq}, 
 and the EIC~\cite{Batell:2022ogj}.

Ref.~\cite{Batell:2022ogj} discusses both lepton number conserving and violating effects of HNLs at the EIC. 
Here we focus on LNV signals and  introduce several new elements compared to   Ref.~\cite{Batell:2022ogj}:
\begin{itemize}
\item First, we perform a more detailed analysis of the  LNV signals of HNLs and corresponding backgrounds at the EIC, 
taking into account parton showering and detector effects. 
\item  Second,  we study  HNL-induced  processes at the EIC that simultaneously violate lepton number 
and change the flavor of the charged lepton, such as $e^- p \to  \ell_\alpha^+ + X$, with $\alpha \in \{\mu, \tau\}$  
(Ref.~\cite{Batell:2022ogj} focused on the  $e^- \to e^+$ channel), 
and discuss possible advantages related to smaller SM backgrounds. 
\item Finally,  we compare the sensitivity of the EIC to that of other colliders (LHC, LEP)  and 
low-energy processes, ranging from $0\nu \beta \beta$ decay to  LFV and LNV  processes 
involving $\mu$ and  $\tau$,  highlighting the discovery potential and complementarity of the EIC in light of these competing searches. 
\end{itemize}

The manuscript is organized as follows: in Section~\ref{sect:setup} we describe 
the theoretical framework. 
In Section~\ref{sect:analysis} we study the expected signal and backgrounds at the EIC, identifying a possible strategy to eliminate the backgrounds.  
We present our results on the sensitivity 
of EIC searches in the context of other experimental probes of HNLs in Section~\ref{sect:pheno}. 
Finally, Section~\ref{sect:conclusion} summarizes the main results and motivate possible future studies.

\section{Theoretical  setup}
\label{sect:setup}

Models with HNLs
 have been extensively studied in the literature (see Ref.\ \cite{Abdullahi:2022jlv} and references therein). 
We explore the EIC sensitivity to HNLs within the following  theoretical framework: 
\begin{itemize}
\item We assume the existence of   $m$ HNLs, of which $n \leq m$ have masses around or below the electroweak (EW) scale, while the remaining $m-n$  have masses well above the EW scale.
 \item We remain agnostic about the underlying model and treat the interactions of the $n$ HNLs
 with masses close to or below the EW scale in the   $\nu$SMEFT~\cite{delAguila:2008ir} framework. 
 The heavier  $m-n$ HNLs  and  possibly other heavy particles  are integrated out, 
and their effect is  encoded in SMEFT operators (built from SM fields) and $\nu$SMEFT operators (involving at least one HNL field)
of dimension  5, 6, 7, ..., so that the  relevant effective Lagrangian takes the form~\cite{Liao:2016qyd} 
(we suppress generation indices to avoid clutter) 
\begin{align}\label{eq:smeft}
	\mathcal L &=  \mathcal L_{SM} + \bar \nu_R\, i\slashed{\partial}\nu_R+ \Big[- \frac{1}{2} \overline {\nu^c_{R}} \,\bar M_R \nu_{R} -\bar L \tilde H Y_\nu \nu_R \nn \\
	& +  \ep_{kl}\ep_{mn}(L_k^T\, C^{( 5)}\,CL_m )H_l H_n \ - 	\overline{\nu^c_{R}} \,\bar M_R^{(5)} \nu_{R} H^\dagger H  + C_{\nu B,\, R}\, \overline {\nu^c_{R}} \sigma^{\mu \nu} \nu_R\,  B_{\mu \nu} +{\rm h.c.}\Big]
	\nn \\
		& +  \mathcal L^{(6)}_{\nu_L}  +  \mathcal L^{(6)}_{\nu_R}
		+  \mathcal L^{(7)}_{\nu_L}  +  \mathcal L^{(7)}_{\nu_R} \,,
\end{align}
where $\mathcal L_{SM}$  is the SM Lagrangian and the three lines in the equation above 
contain operators of dimension four, five, and six / seven, respectively.  
We use the notation  $\Psi^c = C \bar \Psi^T$ for a spinor field $\Psi$ in terms of  $C = - C^{-1} = -C^T = - C^\dagger$, the charge conjugation matrix. 
$L=(\nu_L,\, e_L)^T$ is the left-handed  lepton doublet, $B_{\mu\nu}$ is the  fieldstrength of the hypercharge gauge field, while  $\tilde H = i \tau_2 H^*$ with $H$ denoting the Higgs doublet.
$\nu_{R}$ is a column vector of $n$ HNLs in the `flavor' basis. 
The new dimension-three and -four terms involve  the complex symmetric  $n \times n$ matrix  $\bar M_R$   Majorana mass matrix for the HNLs, and 
the  $3\times n$ Yukawa matrix   $Y_\nu$ that couples the HNLs to the SM fields.  
The dimension-five terms generate Majorana masses for $\nu_L$~\cite{Weinberg:1979sa}  and $\nu_R$~\cite{Aparici:2009fh},
and a transition magnetic moment for $\nu_R$. 
$C^{(5)}$, $\bar M_R^{(5)}$, and $C_{\nu B,\, R}$ are dimensionful Wilson coefficients containing factors of the EFT breakdown scale  $\Lambda$.  
The coefficient of an operator of dimension $d$  scales as $C^{(d)}\sim \Lambda^{4-d}$. 
The  dimension-six  and  -seven effective Lagrangians $\mathcal L^{(6),(7)}_{\nu_L}$ involving only SM fields were  constructed in Refs.~\cite{Grzadkowski:2010es} and \cite{Lehman:2014jma}, respectively. 
Similarly, effective Lagrangians $\mathcal L^{(6),(7)}_{\nu_R}$ contain the complete set of dimension-six and -seven operators involving at least one $\nu_R$ field and were derived in Refs.~\cite{delAguila:2008ir,Liao:2016qyd}.

\item Mixing: we work in the basis where the charged leptons $e^i_{L,R}$ and the quarks $u^i_{L,R}$, $d^i_R$ are mass eigenstates ($i=1,2,3$). After EW symmetry breaking, this means that 
$d^{i,\,\rm gauge}_L = V^{ij} d_L^{j,\,\rm mass}$, 
where $V$ is the Cabibbo-Kobayashi-Maskawa (CKM) matrix. Up to dimension six, 
the mass terms of the neutrinos after EW symmetry breaking but before mass-diagonalization take the form,
  \bea
 \mathcal L_m &=& -\frac{1}{2} \bar {\mathcal N}^c M_\nu \mathcal N +{\rm H.c.}\,,\qquad M_\nu = \bma M_L &M_D^*\\M_D^\dagger&M_R^\dagger \ema \,,\nn\\
M_L &=& -v^2 C^{(5)}\,,\qquad 
M_R = \bar M_R + v^2 \bar M_R^{(5)}\,,\qquad
M_D =\frac{v}{\sqrt{2}} \left[Y_\nu -\frac{v^2}{2}C_{L\nu H}^{(6)}\right]\,.
\eea 
Here $\mathcal N = (\nu_L,\, \nu_R^c)^T$ making $M_\nu$ a $(3+n)\times (3+n)$ symmetric matrix. This matrix can be diagonalized by a  transformation involving the unitary matrix $U$, given by
\bea\label{Mdiag}
U^T M_\nu U =m_\nu = {\rm diag}(m_1,\dots , m_{3+n})\,, \quad\text{with}\quad \mathcal N = U \mathcal N_m\,.
\eea
The  mass eigenstates can be written as 
$\nu = \mathcal N_m+\mathcal N_m^c=\nu^c$, 
and we denote the individual mass basis fields as  
$\nu^T = (\nu_1, \nu_2, \nu_3, N_4, ..., N_{3+n})$.
The first three states $\nu_{1,2,3}$ approximately correspond to active neutrinos with  
 sub-eV masses~\cite{Aker:2019uuj}.
We  denote the mixing between active, lepton-flavor states $\alpha\in\{e,\mu,\tau\}$ with (light and heavy) neutrino mass eigenstates $j \in \{1,\dots , 3+n\}$ by $U_{\alpha j}$. From oscillation data, $\vert U_{\alpha j}\vert\sim\mathcal{O}(1)$ for $j=1,2,3$, while from various constraints (as described below) $\vert U_{\alpha j}\vert\ll1$ for $j\geq4$. 
The charged-current weak  vertices relevant for our analysis read 
\beq
{\cal L}_{CC} =   - \frac{g}{\sqrt{2}} \, W^-_\mu   \sum_{\alpha=e, \mu, \tau}  \, 
\Big( 
\sum_{i=1}^{3}  \,  U_{\alpha i}  \, \bar e^\alpha_L \,   \gamma^\mu  \, P_L \, \nu_i
+ 
\sum_{j=4}^{3+n}  \,  U_{\alpha j}  \, \bar e^\alpha_L \,   \gamma^\mu  \, P_L \, N_j \Big)~. 
\label{eq:LCC}
\eeq
Note that the  $3 \times 3$  block  $U_{\alpha i}$ ($\alpha \in \{e, \mu, \tau \}$ and $i= 1,2,3$)  of the mixing matrix describing active neutrino mixing 
is not unitary.  In this study, we take a phenomenological approach and consider the masses $m_j$ and the matrix elements  $U_{\alpha j}$, with $j>3$ and $\alpha \in \{e, \mu, \tau \}$ 
as independent parameters, focusing on the effects due to the $j=4$ HNL.
Finally, for future reference, we also note that 
 the mixing  $U_{si}$   between ``sterile'' flavors $s = \{1,\dots n\}$ and  mass eigenstates $j = \{1,\dots , 3+n\}$ 
satisfies, by unitarity, $\vert U_{sj}\vert\ll1$ for $j=1,2,3$, while
$\vert U_{sj}\vert$ can be $\mathcal{O}(1)$ for $j\geq4$.

\item   We consider an effective `$3+1$' setup in which a single HNL, $N_4$, with mass $m_4$, is kinematically accessible at the EIC, while the remaining $m-1$ HNLs are heavier, with masses not far above the electroweak scale.

\item  To study the reach of the EIC, we focus on the dimension-4  Yukawa interactions and the mass terms in Eq.~\eqref{eq:smeft}. 
These  generate mixing of heavy and light neutrinos, thus inducing the  $W \ell_\alpha N_j$ charged-current  vertices proportional to the mixing angle $U_{\alpha j}$, 
given in Eq.~\eqref{eq:LCC}, 
which in turn control the  resonant production  of $N_4$ and  its decay,  as shown in Fig.~\ref{fig:signal_feynman}.

\end{itemize}

We note that  in ($\nu$)SMEFT  up to dimension seven, there are a handful of additional operators that, in the resonant production of  $N_4$, can lead to 
an interference effect with the  `mixing' diagrams in Fig.~\ref{fig:signal_feynman}, 
unsuppressed by additional small parameters (charged lepton and quark Yukawa couplings,  
active-sterile mixing  $U_{\alpha j}$ with $j=4,..., 3+n$, and sterile-active 
mixing $U_{si}$ with $i=1,2,3$).  
Their effect can compete with the mixing mechanism if  $U_{\alpha i} \sim (v/\Lambda)^2$ or $U_{\alpha i} \sim (v/\Lambda)^3$, 
which can be realized in models with a relatively low LNV scale. 
 The relevant operators are the dimension-six dipole $\mathcal{O}_{\nu W} = (\bar{L}\sigma_{\mu\nu}\nu_R)\tau^I\tilde{H}W^{I\mu\nu}$~\cite{Liao:2016qyd} and the dimension-seven operators 
 $\mathcal{O}_{NL1} = i\epsilon_{ij} (\overline{\nu_R^c} \gamma_\mu L^i)   ( D^\mu H)^j (H^\dagger H)$ 
 (which produces a shift to the mixing-induced $W \ell \nu_R$ vertex)  and 
 $\mathcal{O}_{QNQLH2}= \epsilon_{ij} (\bar Q \nu_R) (\overline{Q^{c,i}} L^j)  H$~\cite{Liao:2016qyd}~\footnote{In 
 Section~\ref{sect:pheno} we will present a concise discussion of how similar dimension-six and dimension-seven 
 operators affect other observables that are used to constrain the mixing parameters $U_{\alpha 4}.$}. 
 Other dimension-six and seven operators can also contribute to resonant production, but, as they do not interfere with the mixing mechanism, they do not affect the bounds on $U_{\alpha 4}$ we will present in Section \ref{sect:pheno}.
Here,  we focus  on the mixing mechanism to explore the relative competitiveness of the EIC,
 leaving the more general analysis for future work. We expect that the  results 
 based on the mixing mechanism, summarized in Section~\ref{sect:results1}, 
 will provide a reasonable indication of the competitiveness and complementarity  of the EIC with measurements 
 at LEP and the LHC.

In the limit $n=1$ and ignoring the higher dimensional $(\nu)$SMEFT operators 
($\Lambda \gg v$), we obtain a more constrained model, in which  
many low-energy processes  (such as $0 \nu \beta \beta$ decay, lepton-flavor violating 
processes such as $\mu \to e \gamma$,  
leptonic and semi-leptonic  weak  decays of mesons, etc.) 
indirectly  probe  the same HNL parameters ($U_{\ell 4}$ and $m_4$)  controlling the EIC signal. 
Variants of this setup have been studied in detail in Refs.~\cite{Atre:2009rg,Drewes:2015iva,Abada:2017jjx,Bolton:2019pcu}. 
In Appendix \ref{app:details} we discuss  updates to these analyses relevant for our  study. 
The  constraints obtained in the  limit $n=1$  ignoring $(\nu)$SMEFT operators  will be presented 
in Section~\ref{sect:results2}.

\section{EIC analysis}
\label{sect:analysis}

Collider searches for HNLs in $ep$ collisions have been proposed in the context of HERA~\cite{Almeida:1990vt,Buchmuller:1991tu,Buchmuller:1992wm,Ingelman:1993ve}\,\footnote{The H1 Collaboration also has searches for excited neutrino states at HERA \cite{H1:2008qoo}.}, the LHeC~\cite{Blaksley:2011ey,Duarte:2014zea,Mattelaer:2016ynf,Antusch:2016ejd}, and more recently the EIC~\cite{Batell:2022ogj}. As in many of them, the event simulation at the matrix-element (ME) level is performed by MadGraph (\texttt{MadGraph5\_aMC v3.6.2}) \cite{Alwall:2014hca}, while the parton shower (PS) and hadronization are handled by Pythia (\texttt{Pythia8 v8.315}) \cite{Bierlich:2022pfr}. The `SM + HNLs' model realization in MadGraph was directly taken from the FeynRules (\texttt{FeynRules v2.3})~\cite{Alloul:2013bka} model database, based on the implementation by the authors of Refs. \cite{Alva:2014gxa,Degrande:2016aje}. 

From the technical capabilities described in the EIC Yellow Report \cite{EIC:YellowReport}, we select the following operation point for our analysis: center of mass energy $\sqrt{s}=141\,{\rm GeV}$ ($E_e\times E_p=18\,{\rm GeV}\times275\,{\rm GeV}$), integrated luminosity $\mathcal{L}=100\,{\rm fb}^{-1}$, and electron polarization $P_e=70\%$. The detector response is managed by Delphes (\texttt{Delphes3 v3.5.0}) \cite{deFavereau:2013fsa} using a customized card \cite{Arratia:2020azl,Arratia:2021uqr} implementing the specified parameters in the EIC Yellow Report \cite{EIC:YellowReport}. The same report shows that there are no specific detector capabilities for muons, which are essential for our flavorful analysis. In the spirit of the discussion of a second detector \cite{EIC:2ndDetector2023, EIC:2ndDetector2024, EIC:2ndDetector2025}, we have modified the aforementioned Delphes card to include a future, hypothetical muon detector with the same technical capabilities (efficiency, angular response, energy dependence, etc.) as for electrons.

Multi-jet events generated at fixed-order in QCD may suffer from overestimated cross sections due to 
large logarithms $\sigma(e^-p\to N_4\ell^{\pm}+k\,j)\sim\log(\hat s/p_{Tj}^2)$ \cite{Degrande:2016aje}, so we impose a set of basic generator-level $p_{Tj}$ cuts ($p_{Tj}>10\,{\rm GeV}$), for both signal and background, to ensure the perturbativity of these logarithms \cite{D0:2000fnp}. We additionally implement some complementary generator-level cuts ($p_{T\ell}>5\,{\rm GeV}$, $|\eta_{j,\ell}|<5$, $\Delta R_{jj,\,j\ell}>0.4$) and perform parton showering to properly populate the soft region of the $p_{Tj}$ spectrum\,\footnote{Here, $\Delta R_{ab} \equiv \sqrt{(\Delta \eta_{ab})^2 + (\Delta \phi_{ab})^2}$ denotes the separation between objects $a$ and $b$ in the pseudorapidity–azimuthal angle plane. In particular, $\Delta R_{jj}$ refers to the jet–jet separation and $\Delta R_{j\ell}$ to the jet–lepton separation.}. To avoid double-counting between matrix-elements and shower emissions, we employ the MLM matching and merging scheme \cite{Pascoli:2018heg}, as implemented in MadGraph and Pythia, with \texttt{xqcut=10}.\\

\subsection{Signal}
\begin{figure}
\begin{center} 
\includegraphics[width=0.70\textwidth]{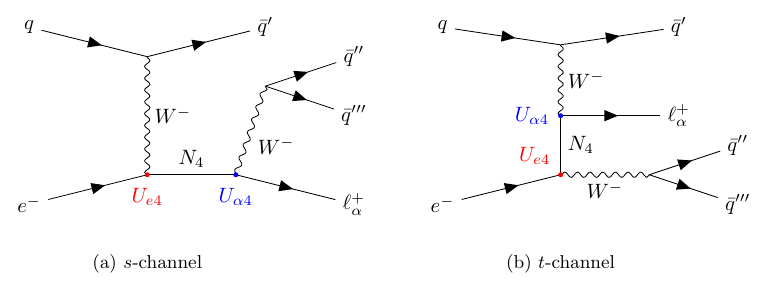}
\end{center} 
\vspace*{-1.cm}
\caption{Parton-level diagrams for the process $e^-p\to \ell_\alpha^++k\,j+X$ ($\alpha\in\{e,\mu,\tau\}$), where $k$ denotes the number of jets. 
}
\label{fig:signal_feynman}
\end{figure}

Based on the experience collected in the existing literature, we study the LNV signal $e^-p\to \ell_\alpha^++k\,j+X$ ($\alpha\in\{e,\mu,\tau\}$), where $k$ denotes the number of jets, mediated by the exchange of a sterile neutrino, as depicted in Fig. \ref{fig:signal_feynman} at the parton level. While previous strategies have focused solely on $k=3$ (\emph{e.g.}, Ref. \cite{Batell:2022ogj}), a more inclusive approach with $k=2$ increases the discovery potential \cite{Mattelaer:2016ynf}. They both correspond to the same parton-level event in Fig. \ref{fig:signal_feynman}, where the difference occurs at the reconstruction level. 

With respect to the two contributions depicted in Fig. \ref{fig:signal_feynman}, the $s$-channel dominates for the mass range of interest. Moreover, for the mass range of interest, the width of the sterile neutrino~\cite{Gorbunov:2007ak, Atre:2009rg, Helo:2010cw, Shuve:2016muy, Bondarenko:2018ptm} is small enough (compared to its mass) to 
justify the narrow-width approximation (NWA)\,\footnote{Recent developments beyond the NWA for large-width scenarios can be found, for instance, in Ref.~\cite{Shen:2025nkr}.} for a resonant production \cite{Abada:2022wvh}:
\begin{align}
\sigma(e^-\,p\to\ell_\alpha^++k\,j+X)&\approx\sigma(e^-\,p\to N_4\,j+X)\,\mathrm{Br}(N_4\to\ell_\alpha^++k'\,j)
=\frac{|U_{e4}U_{\alpha4}|^2}{U_4^2}\,\sigma_0(m_4)\,,
\label{eq:sigma_res}
\end{align}
where $\sigma_0(m_4)$ is a function that depends only on the HNL mass, $k'$ is the number of jets produced by the decay of $N_4$, and we have defined the mixing angle combination
\begin{align}
U_4 \equiv \sqrt{|U_{e4}|^2 + |U_{\mu4}|^2 + |U_{\tau4}|^2}\,.
\label{eq:def_U4}
\end{align}
Note  that the width of $N_4$, $\Gamma_{N_4}\propto U_4^2$, is dominated by the weak channels $N_4\to Z^{(\ast)}\nu_i$ ($i=1,2,3$) and $N_4\to W^\pm{}^{(\ast)}\ell^{\mp}_\alpha$ ($\alpha\in\{e,\mu,\tau\}$) for $10\,{\rm GeV}\lesssim  m_4<100\,{\rm GeV}$ \cite{Bondarenko:2018ptm}. Eq. \eqref{eq:sigma_res} effectively shows how the signal cross section can be factorized into a coupling-dependent term and a mass-dependent piece, simplifying the analysis for a parameter scan.

The factorization in Eq. \eqref{eq:sigma_res} allows us to study the signal sensitivity in terms of the HNL mass $(m_4)$ and the coupling combination $|U_{e4}U_{\alpha 4}|^2/U_4^2$ for the three different channels ($\alpha\in\{e,\mu,\tau\}$). For the electron channel, namely $e^-p\to e^+jj(j)$, we target an isolated positron in the final state. For both the muon and tau channels, we opted for the reconstruction of a $\mu^+$. While the $\tau$ decay mode into muons has a branching ratio $\approx 17\%$ \cite{ParticleDataGroup:2024cfk}, it provides a relatively clean environment compared with the hadronic decay modes \cite{Cirigliano:2021img}. Improving the analysis in the hadronic channels could highly impact the EIC
reach, but we defer this dedicated study to future work.

In principle, the same strategy could also be applied to the lepton-number conserving (LNC) counterpart of the process for all three lepton-flavor channels, since the production mechanism and resonant kinematics remain unchanged. Particularly, in the LNC case for the electron channel, the corresponding SM backgrounds are significantly larger, substantially reducing the achievable sensitivity. In the muon and $\tau$ channels, we expect the LNC and LNV channels to have similar sensitivities.

\subsection{Background}

The SM background for LNV searched in $ep$ collisions has been described in the literature \cite{Almeida:1990vt,Buchmuller:1991tu,Buchmuller:1992wm,Ingelman:1993ve,Blaksley:2011ey,Duarte:2014zea,Mattelaer:2016ynf,Antusch:2016ejd,Batell:2022ogj} and can be categorized in three types:

\begin{itemize}
\item \textbf{Deep inelastic scattering (DIS)}\\
In neutral-current DIS (NC-DIS), $e^-p\to e^-+X$, the scattered lepton yields an isolated electron signature, while the multi-jet component originates from QCD radiation. In the electron channel, an important reducible background arises from charge misidentification, whereby isolated electrons are reconstructed as positrons, primarily due to hard bremsstrahlung in the tracker material\,\footnote{The so-called trident process $e^-\to e^-\gamma^\ast\to e^-e^+e^-$~\cite{ATLAS:2017xqs,Muskinja:2643902,ATLAS:2019qmc}. Since the bremsstrahlung rate scales approximately as $m_\ell^{-4}$~\cite{Jackson:1998nia}, this effect is negligible for muons.}. Following analogous studies in the LHC context \cite{ATLAS:2017xqs,Muskinja:2643902,ATLAS:2019qmc,Harz:2021psp}, we model this contribution through an electron charge-misidentification probability $P_{\rm misID}$. While a realistic treatment would require a detector-dependent parametrization (\emph{e.g.}, angular and energy dependence), we adopt a constant benchmark value $P_{\rm misID}=10^{-3}$ \cite{ATLAS:2019qmc,Batell:2022ogj}, which is conservative and likely underestimates the achievable sensitivity once a full detector study is performed.

In the muon and $\tau$ channels, NC-DIS constitutes an irreducible background, as hadronization can produce secondary decays yielding muons and $\tau$ leptons. We simulate multi-jet DIS samples by matching and merging matrix-element configurations with one, two, and three parton-level jets using MadGraph interfaced with Pythia.

Our simulations show that NC-DIS is one of the main background components across all lepton-flavor channels, in agreement with the results of Ref.~\cite{Batell:2022ogj} for the electron/positron channel. 

In charged-current DIS (CC-DIS), $e^-p\to \nu_e+X$, the absence of a prompt charged lepton at the hard-scattering level does not preclude lepton production at later stages. Hadronization and subsequent hadron decays can generate secondary electrons, muons, and $\tau$ leptons, thereby populating all three signal channels. As a result, CC-DIS constitutes an irreducible background for the electron, muon, and $\tau$ channels.
Our simulations indicate that CC-DIS represents a sizable contribution to the total background.

\item \textbf{Heavy quark production}\\
Boson-gluon fusion into a pair of heavy quarks ($e^-p\to e^-+\bar QQ+X$) may contribute as a background source due to the presence of isolated leptons in the decay of the heavy quarks. The scattered electron is not reconstructed, either because it is collimated along the beamline or because it participates in a charged-current interaction producing a neutrino and a $W$ boson\,\footnote{In principle, charge misidentification could also play a role, but the effect is negligible.}. We simulated heavy-quark-pair production (matched and merged up to one extra jet) for $b$-quarks, keeping the default MadGraph settings to treat $\{u,d,s,c\}$ as light quark flavors. Dealing with charm quarks also as heavy flavors, although highly motivated in the context of the EIC \cite{Arratia:2020azl,Li:2025lxr}, goes beyond the scope of this work. 

We find that this background is sizable for all lepton-flavor channels. Ref.~\cite{Batell:2022ogj} did not include this channel in their background estimation for the electron/positron channel. 
Since the analysis of Ref.~\cite{Batell:2022ogj} was carried out at the parton level, \emph{i.e.}, neglecting hadronization effects, the omission of heavy-quark–pair production may be attributable to its incompatibility with this level of description.

\item \textbf{$\boldsymbol{W}$ boson production}\\
Similar to the previous case, the production of a $W$ boson and its subsequent decay will contribute with either jets or leptons. The main contributions to this background source come from $e^-p\to \nu_e+W^\pm +X$ and $\gamma p\to W^\pm +X$, where the photon is radiated by the incoming electron. Subsequent $W$ boson decays are responsible for the presence of isolated leptons. In MadGraph, the photoproduction is simulated using the Weizsäcker-Williams approximation \cite{Frixione:1993yw}. 

Our results indicate that this background class is negligible compared to the previous two, in agreement with the findings of Ref. \cite{Batell:2022ogj}. We therefore neglect this contribution in the subsequent analysis.
\end{itemize}

\subsection{Sensitivity}
\label{sec:EIC_analysis_sensitivity}
To evaluate the 
EIC sensitivity to HNLs, we perform a proof-of-principle cut-based analysis. Although a more sophisticated approach involving multivariate analysis techniques or artificial intelligence/machine learning (AI/ML) could yield better performance, we emphasize the goal of this work is to motivate future developments (both theoretical and experimental) to improve the EIC's sensitivity for BSM physics.

We generated $10^7-10^8$ events for each of the relevant background types (heavy-quark production, CC-DIS, and NC-DIS), and $10^6$ events for the signal process $e^-p\to \ell_\alpha^++k\,j+X$ ($\alpha\in\{e,\mu,\tau\}$) with fixed HNL masses
$m_4=\{10,25,40,55,70,85,100\}\,{\rm GeV}$. To identify the relevant signal-like events, we impose a preliminary set of signal selection cuts listed in Box \ref{box:signal_selection}.

\begin{tcolorbox}[title=\textbf{Box \refstepcounter{cutbox}\thecutbox\label{box:signal_selection}}. Signal selection cut (Cut-SS),colback=white,opacityback=100,breakable]
\begin{itemize}
\item Exactly two or three jets: $N_j=2,3$.
\item Exactly one positively charged lepton:
\begin{itemize}
\item For the electron channel, $N_{e^+} = 1$.
\item For the muon and tau channels, $N_{\mu^+} = 1$, with an electron veto applied.
\item For backgrounds involving charge misidentification, we select one electron $(N_{e^-}=1)$ and weight the events by the charge misidentification probability $P_{\rm misID}$.
\end{itemize}
\item Jet–jet and jet–lepton isolation, $\Delta R_{jj,\,j\ell} > 0.4$, and detector acceptance requirement $|\eta_{j,\ell}| < 5$.
\end{itemize}
\end{tcolorbox}

\begin{figure}
\begin{center} 
\includegraphics[width=1.0\textwidth]{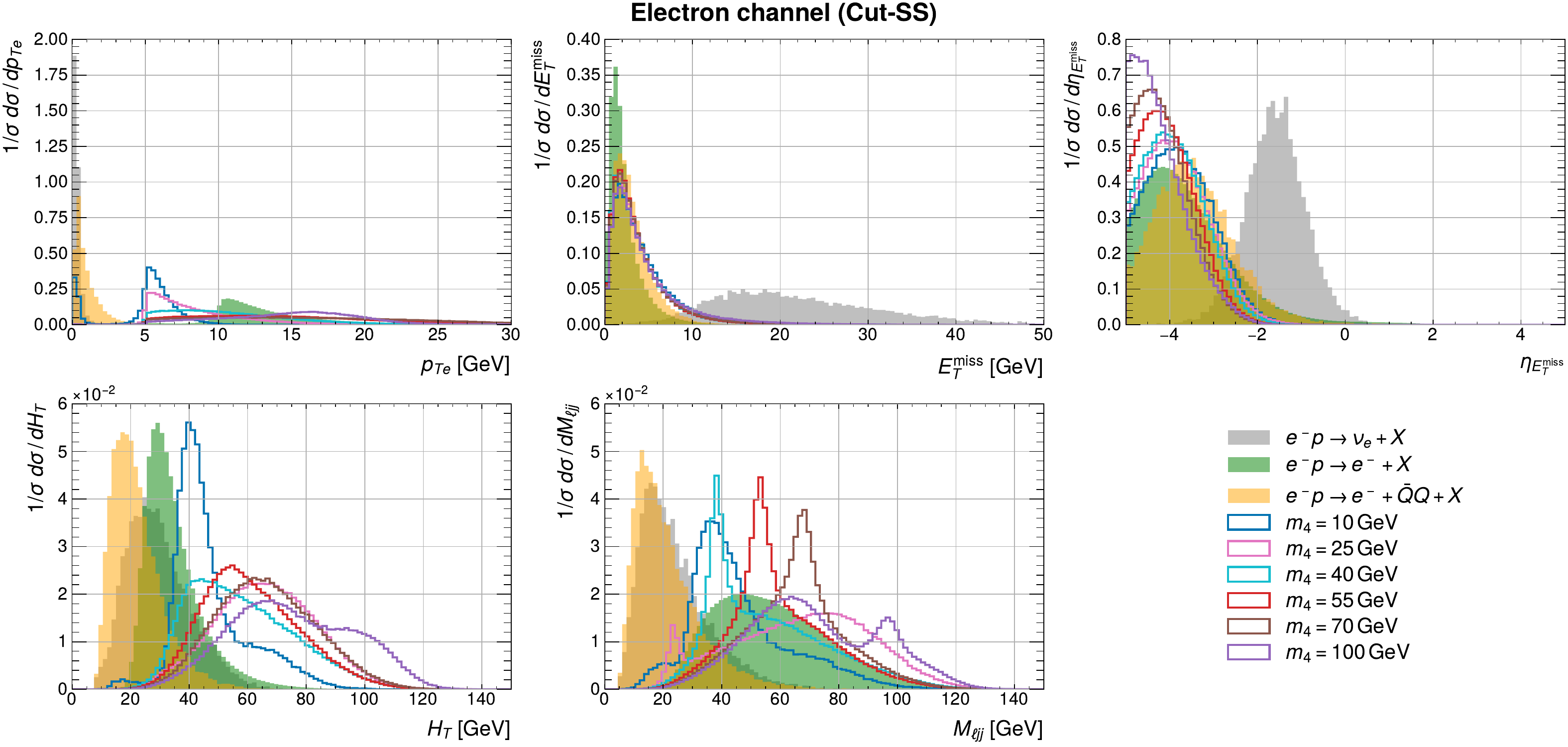}
\end{center} 
\caption{Kinematic distributions for the electron channel $e^-p\to e^+jj(j)$ after imposing the signal selection cuts (Cut-SS) in Box~\ref{box:signal_selection}. Background contributions are shown as filled histograms, while representative HNL mass benchmarks are displayed as solid lines.}
\label{fig:distributions_electron_bkg}
\end{figure}

\begin{figure}
\begin{center} 
\includegraphics[width=1.0\textwidth]{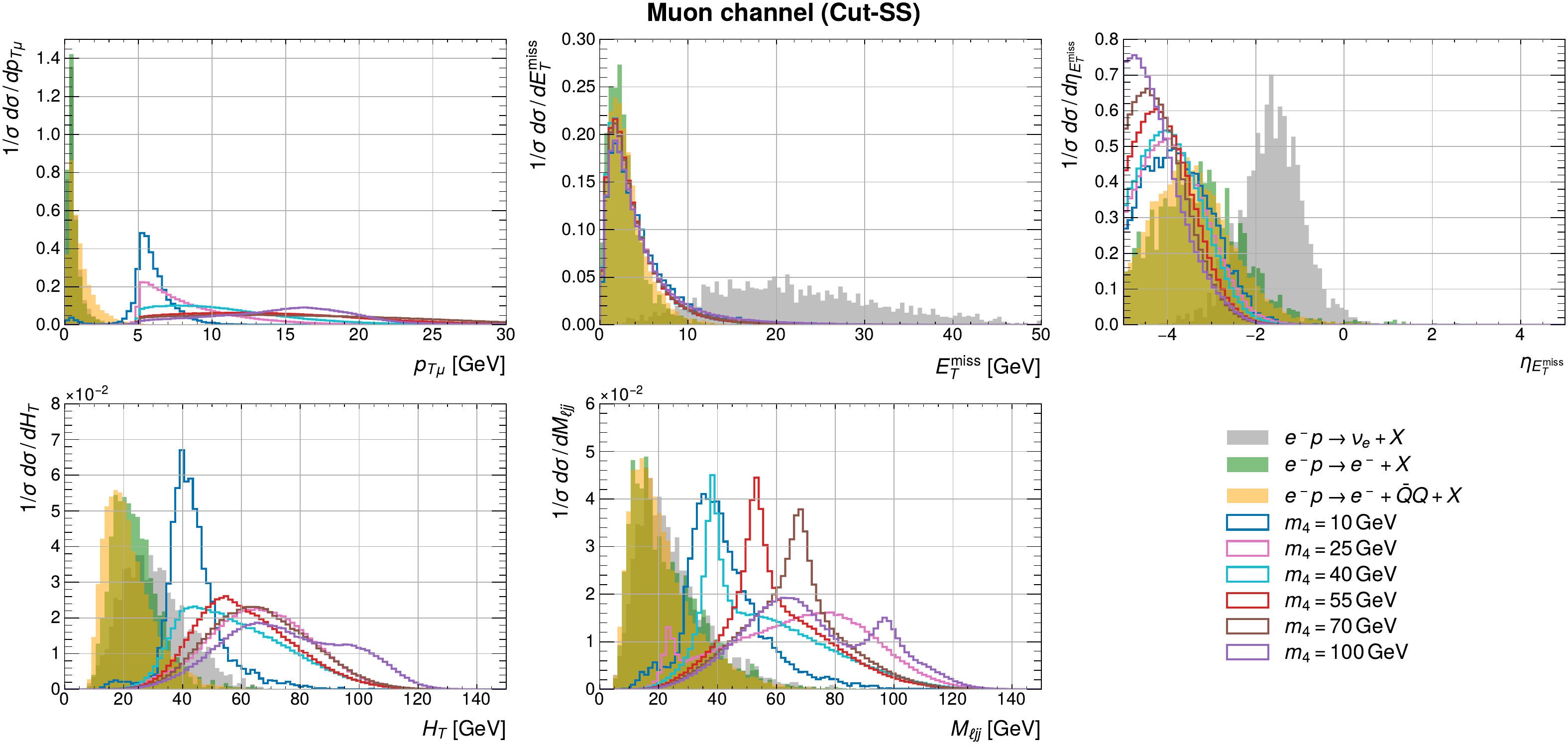}
\end{center} 
\caption{Kinematic distributions for the muon channel $e^-p\to\mu^+jj(j)$, with the same conventions as in Fig.~\ref{fig:distributions_electron_bkg}.}
\label{fig:distributions_muon_bkg}
\end{figure}

\begin{figure}
\begin{center} 
\includegraphics[width=1.0\textwidth]{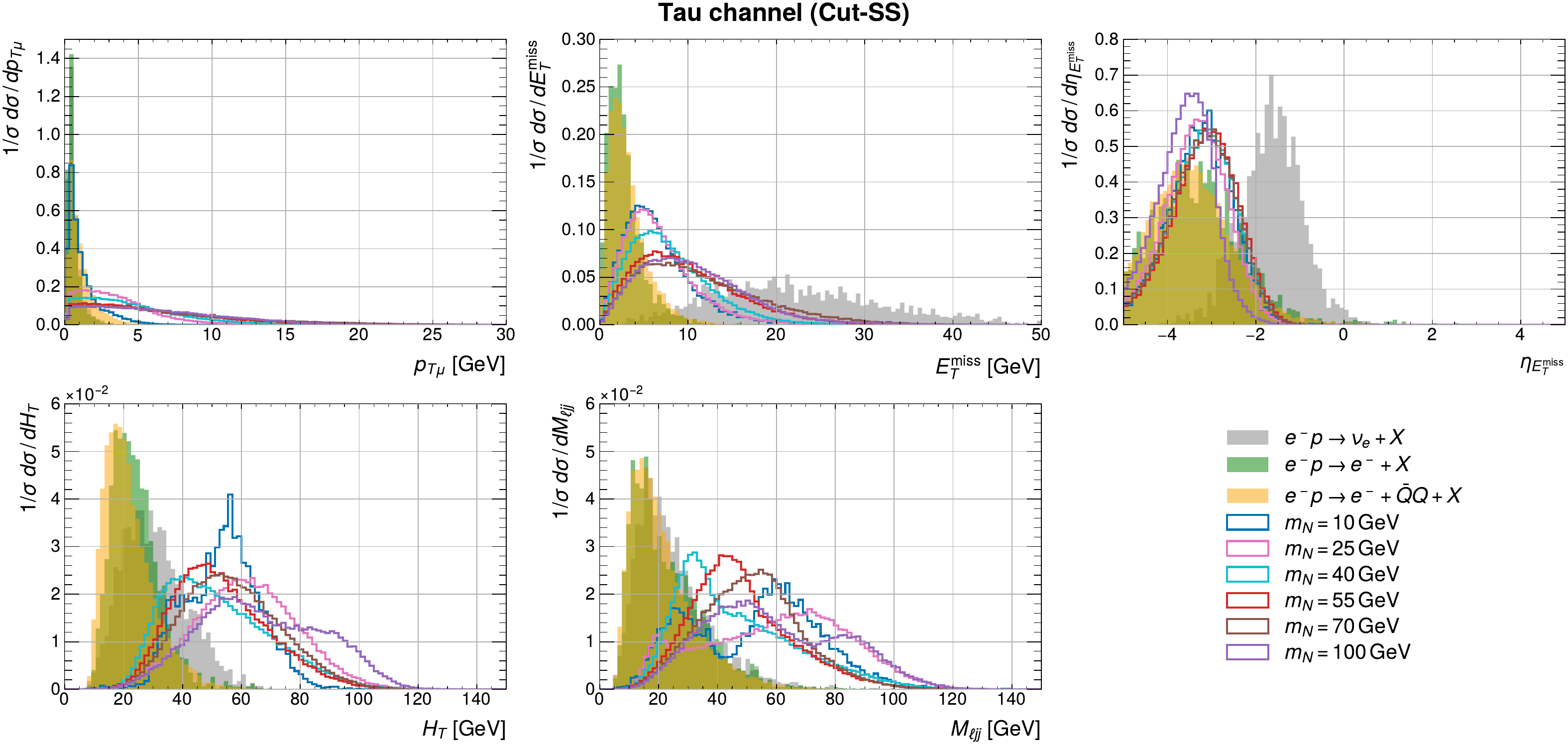}
\end{center} 
\caption{Kinematic distributions for the $\tau$ channel $e^-p\to\tau^+jj(j)$, with the same conventions as in Fig.~\ref{fig:distributions_electron_bkg}.}
\label{fig:distributions_tau_bkg}
\end{figure}

To identify a feasible set of kinematic cuts that isolate signal from background, we analyzed the distributions of several common kinematic variables after applying the signal selection criteria (Cut-SS) in Box~\ref{box:signal_selection}. The most relevant results are displayed in Figs. \ref{fig:distributions_electron_bkg}, \ref{fig:distributions_muon_bkg}, and \ref{fig:distributions_tau_bkg} for the electron, muon, and $\tau$ channels, respectively. For many kinematic variables (for example, azimuthal angle or pseudorapidity distributions for jets and leptons), there is significant overlap between the background and signal. The transverse momentum of the final-state lepton, $p_{T\ell}$, the missing transverse energy, $E_T^{\rm miss}$, and the transverse momenta of the leading jets, $p_{Tj}^{(i)}$ $(i=1,2,3)$, provide good discrimination between signal and background. We therefore characterize the jet activity through $H_T \equiv \sum_i p_{Tj}^{(i)}$, the scalar sum of the jet transverse momenta. We also include the reconstructed invariant mass $M_{\ell jj}$ of the lepton and any pair of jets, which is commonly used in searches involving direct production \cite{Almeida:1990vt,Buchmuller:1991tu,Buchmuller:1992wm,Ingelman:1993ve,Blaksley:2011ey,Duarte:2014zea,Mattelaer:2016ynf,Antusch:2016ejd,Batell:2022ogj}.

Our results differ from similar electron-channel distributions reported in Ref.~\cite{Batell:2022ogj} due to two main effects. First, the generator-level cuts imposed in our analysis (introduced to ensure the stability and reliability of the matrix-element calculation) are more restrictive than those applied in Ref.~\cite{Batell:2022ogj}. These tighter requirements already modify the event kinematics at the parton level, particularly in the soft and low-mass regions. Second, our simulation includes parton showering and hadronization, which further distorts the kinematic distributions of jets and leptons, whereas Ref.~\cite{Batell:2022ogj} is based on a purely parton-level study. The combined impact of these effects is most clearly visible in the $M_{\ell jj}$ distributions shown in Fig.~\ref{fig:distributions_electron_bkg}. Although resonant structures around $M_{\ell jj} \sim m_4$ are present for all HNL masses considered, the peaks are significantly broadened relative to the parton-level results of Ref.~\cite{Batell:2022ogj}. Consequently, the mass-window variable $\Delta M^{\rm min}\equiv \underset{a,b}{\min}(|M_{\ell j_a j_b}-m_4|)$, which was adopted in Ref.~\cite{Batell:2022ogj}, does not provide the same level of discrimination in our setup. For this reason, we do not include $\Delta M^{\rm min}$ among the observables used in our analysis.

As shown in Figs.~\ref{fig:distributions_electron_bkg}--\ref{fig:distributions_tau_bkg}, requiring $H_T \gtrsim 50\,{\rm GeV}$ provides strong separation between signal and background, particularly in the muon and $\tau$ channels. In addition, imposing $p_{T\ell} \gtrsim 5\,{\rm GeV}$ efficiently suppresses the heavy-quark production background in all channels and further reduces the NC-DIS contribution in the muon and $\tau$ modes. In the $\tau$ channel, the $p_{T\ell}$ distribution of the signal is shifted toward lower values, since the isolated muon arises from the decay of a $\tau$, leading to a softer transverse-momentum spectrum. Hadronization effects and detector resolution induce a substantial overlap between the $E_T^{\rm miss}$ spectra of signal and background. An exception arises in the $\tau$ channel, where additional neutrinos from leptonic $\tau$ decays broaden the signal $E_T^{\rm miss}$ distribution, motivating the requirement $E_T^{\rm miss} \gtrsim 5\,\mathrm{GeV}$. The CC-DIS background also displays a broader and more forward $E_T^{\rm miss}$ distribution, which could in principle motivate additional upper bounds, $E_T^{\rm miss} \lesssim 20\,\mathrm{GeV}$ and $\eta_{E_T^{\rm miss}} \lesssim -2$. In practice, however, the existing $H_T$ and $p_{T\ell}$ requirements already suppress the CC-DIS contribution efficiently, rendering further cuts redundant.

Consolidating these results, we propose a set of kinematic cuts for each lepton-flavor channel, defined explicitly in Boxes~\ref{box:cut_e}--\ref{box:cut_tau}. Table~\ref{tab:cut-flow} summarizes their performance for two representative HNL mass benchmarks, illustrating how the successive selections suppress the dominant backgrounds while retaining signal efficiency. Entries expressed as $\mathcal{O}(\cdot)$ indicate that no simulated events survive the corresponding cut combination; the quoted value reflects the upper bound obtained from a single event with unit weight in the Monte Carlo sample.

\begin{tcolorbox}[title=\textbf{Box \refstepcounter{cutbox}\thecutbox\label{box:cut_e}}. Electron channel cuts (Cut-$e$),colback=white,opacityback=100,breakable]
\begin{itemize}
\item All the requirements from Cut-SS in Box \ref{box:signal_selection}.
\item $p_{Te}>5\,{\rm GeV}$ and $H_T>50\,{\rm GeV}$.
\end{itemize}
\end{tcolorbox}

\begin{tcolorbox}[title=\textbf{Box \refstepcounter{cutbox}\thecutbox\label{box:cut_muon}}. Muon channel cuts (Cut-$\mu$),colback=white,opacityback=100,breakable]
\begin{itemize}
\item All the requirements from Cut-SS in Box \ref{box:signal_selection}.
\item $p_{T\mu}>5\,{\rm GeV}$ and $H_T>50\,{\rm GeV}$.
\end{itemize}
\end{tcolorbox}

\begin{tcolorbox}[title=\textbf{Box \refstepcounter{cutbox}\thecutbox\label{box:cut_tau}}. $\tau$ channel cuts (Cut-$\tau$),colback=white,opacityback=100,breakable]
\begin{itemize}
\item All the requirements from Cut-SS in Box \ref{box:signal_selection}.
\item $p_{T\mu}>5\,{\rm GeV}$, $H_T>50\,{\rm GeV}$, and $E_{T}^{\rm miss}>5\,{\rm GeV}$.
\end{itemize}
\end{tcolorbox}


\begin{table}
\centering
\begin{tabular}{llrrrrr}
\toprule
  \multicolumn{2}{l}{\multirow[c]{3}{*}{\makecell[l]{\textbf{Lepton} \\ \textbf{flavor} \\ \textbf{channel}}}} & \multicolumn{3}{c}{\textbf{Background}} & \multicolumn{2}{c}{\textbf{Mass benchmark}} \\
\cmidrule(lr){3-5}\cmidrule(lr){6-7}
  \multicolumn{2}{l}{} & NC-DIS & CC-DIS & Heavy-quark & $m_4=10\,{\rm GeV}$ & $m_4=70\,{\rm GeV}$ \\
  \multicolumn{2}{l}{} & {\footnotesize [pb]} & {\footnotesize [pb]} & {\footnotesize [pb]} & {\footnotesize [pb]} & {\footnotesize [pb]} \\
\midrule
  \multirow{3}{*}{$e$} & ME+PS & $2.37 \times 10^{0}$ & $2.36 \times 10^{1}$ & $3.32 \times 10^{1}$ & $1.01 \times 10^{-2}$ & $7.47 \times 10^{-2}$ \\
   & Cut-SS & $9.15 \times 10^{-2}$ & $5.35 \times 10^{-3}$ & $6.50 \times 10^{-2}$ & $7.99 \times 10^{-5}$ & $3.08 \times 10^{-2}$ \\
   & Cut-$e$ & $7.28 \times 10^{-3}$ & $8.91 \times 10^{-7}$ & $3.54 \times 10^{-6}$ & $7.94 \times 10^{-6}$ & $2.55 \times 10^{-2}$ \\
  \cmidrule(lr){1-7}
  \multirow{3}{*}{$\mu$} & ME+PS & $2.37 \times 10^{3}$ & $2.36 \times 10^{1}$ & $3.32 \times 10^{1}$ & $9.90 \times 10^{-3}$ & $7.47 \times 10^{-2}$ \\
   & Cut-SS & $5.56 \times 10^{-2}$ & $8.45 \times 10^{-4}$ & $6.65 \times 10^{-2}$ & $7.64 \times 10^{-5}$ & $3.77 \times 10^{-2}$ \\
   & Cut-$\mu$ & $<\mathcal{O}(10^{-5})$ & $<\mathcal{O}(10^{-7})$ & $1.07 \times 10^{-5}$ & $9.16 \times 10^{-6}$ & $3.13 \times 10^{-2}$ \\
  \cmidrule(lr){1-7}
  \multirow{3}{*}{$\tau$} & ME+PS & $2.37 \times 10^{3}$ & $2.36 \times 10^{1}$ & $3.32 \times 10^{1}$ & $9.93 \times 10^{-3}$ & $7.47 \times 10^{-2}$ \\
   & Cut-SS & $5.56 \times 10^{-2}$ & $8.45 \times 10^{-4}$ & $6.65 \times 10^{-2}$ & $7.58 \times 10^{-5}$ & $6.13 \times 10^{-3}$ \\
   & Cut-$\tau$ & $<\mathcal{O}(10^{-5})$ & $<\mathcal{O}(10^{-7})$ & $3.55 \times 10^{-6}$ & $5.95 \times 10^{-8}$ & $1.95 \times 10^{-3}$ \\
\bottomrule
\end{tabular}
\caption{
Cut-flow table for the dominant SM backgrounds and the LNV signal in the three lepton-flavor channels. All entries are cross sections in pb. The collider setup assumes $\sqrt{s}=141\,{\rm GeV}$ ($E_e\times E_p=18\,{\rm GeV}\times275\,{\rm GeV}$) and electron polarization $P_e=70\%$. The rows labeled ME+PS correspond to matched matrix-element plus parton-shower simulations prior to any kinematic selections. Cut-SS denotes the signal selection requirements of Box~\ref{box:signal_selection}, while Cut-$e$, Cut-$\mu$, and Cut-$\tau$ refer to the channel-dependent cuts defined in Boxes~\ref{box:cut_e}--\ref{box:cut_tau}. The signal benchmarks correspond to $m_4 = 10\,{\rm GeV}$ and $m_4 = 70\,{\rm GeV}$ with $|U_{e4}|=|U_{\mu4}|=|U_{\tau4}|=1$. When applicable (\emph{e.g.}, in the electron channel), the results have been weighted by a charge-misidentification probability $P_{\rm misID}=10^{-3}$. Entries expressed as $\mathcal{O}(\cdot)$ indicate that no simulated events survive the corresponding cut combination; the quoted value reflects the upper bound obtained from a single event with unit weight in the Monte Carlo sample.}
\label{tab:cut-flow}
\end{table}

\section{Phenomenology} 
\label{sect:pheno}

In this section, we discuss the impact of HNL searches at the  EIC 
in the context of other HNL probes.    
In the mass region of interest ($m_4 \sim$~10-100 GeV), the strongest constraints on HNL couplings 
arise from~\cite{Atre:2009rg,Drewes:2015iva,Abada:2017jjx,Bolton:2019pcu,Blennow:2023mqx,Fernandez-Martinez:2023phj} 
(i) collider searches at the LHC; 
(ii) $Z$ decays at LEP; 
(iii) weak decays of mesons; 
(iv) LFV and (to a lesser extent) LNV decays of leptons and mesons; 
(v) $0\nu \beta \beta$ decay. 
In the context of $3+1$ models without inclusion of $\nu$SMEFT operators, these processes are studied 
together, as they all constrain the  HNL mixing parameters $U_{\alpha 4}$ and $m_4$.

As discussed in Section~\ref{sect:setup},  here we consider a more general setup that includes $(\nu)$SMEFT operators 
up to dimension seven.  In this framework the  EIC resonant cross section $e^- p \to  \ell_\beta^+ +X$ and the corresponding `crossed' process at the LHC, 
$pp \to  e^\pm \ell_\beta^\pm + X$,   
probe the same  couplings, though not necessarily with the same 
relative weight, given the different energies involved. 
Here, we have considered only the mixing-induced production and decay. 
On the other hand,   probes in classes (ii) to (v) above are sensitive to different  combinations of  mixing angles and 
$(\nu)$SMEFT Wilson Coefficients compared to the EIC resonant cross section.

We  summarize here the  $(\nu)$SMEFT contributions to probes (ii) to (v) that can interfere with the 
mixing-induced contributions, possibly diluting the corresponding indirect bounds on  $|U_{\alpha 4} U_{\beta 4}|$.  
As in Section~\ref{sect:setup}, we focus on effects that are suppressed only by 
the $(\nu)$SMEFT expansion parameter $(v/\Lambda)^n$ ($n=2,3$), without  
additional suppression due to ``wrong chirality" in the neutrino, charged leptons, and quarks.  
(ii)  The process $Z \to N_4 \nu_i$ ($i=1,2,3$) receives  contributions from 
a linear combination of the dimension-six operators 
$\mathcal{O}_{\nu B} = (\bar{L}\sigma_{\mu\nu}\nu_R ) \tilde{H}B^{\mu\nu}$~\cite{Liao:2016qyd} 
and 
$\mathcal{O}_{\nu W} = (\bar{L}\sigma_{\mu\nu}\nu_R )\tau^I\tilde{H}W^{I\mu\nu}$~\cite{Liao:2016qyd}, 
as well as the  operators 
$\mathcal{O}_{NL1} = i\epsilon_{ij} (\overline{\nu_R^c} \gamma_\mu L^i)   ( D^\mu H)^j (H^\dagger H)$  and 
$\mathcal{O}_{NL2} = \epsilon_{ij} ( \overline{\nu_R^c} \gamma_\mu L^i) H^j   
(i  H^\dagger \overleftrightarrow{D_\mu} H)$ \cite{Liao:2016qyd} arising at dimension seven. 
(iii)    Weak decays $d_j \to u_i \ell_\alpha \bar \nu_\alpha$  are affected by heavy-light neutrino mixing and by 
the dimension-six operators in  SMEFT, resulting in effective vertices 
$G_F V_{u_i d_j}  \times [1  + (1/2) (   |U_{e4}|^2 + |U_{\mu 4}|^2 - |U_{\alpha 4}|^2 )   + \epsilon_L^{(\alpha)} ]$, 
where  $ \epsilon_L^{(\alpha)} $  denotes  a linear combination of  dimension-six SMEFT couplings~\cite{Cirigliano:2009wk,
Gonzalez-Alonso:2017iyc}.
(iv) LFV  decays of charged leptons and mesons can receive interfering contributions   
from $N_{5}-N_{3+n}$ as well as from dimension-six SMEFT operators. 
(v) Similarly,   in addition to the combination  $U_{e4}^2/m_4$,  $0\nu \beta \beta$ decay  
receives interfering contributions from  $N_{5}-N_{3+n}$ and from  
dimension-seven SMEFT operators~\cite{Lehman:2014jma,Cirigliano:2017djv}.

Based on the above discussion, we present  our results as follows:
\begin{itemize}
\item  First, we show the EIC sensitivity to the combination $|U_{\alpha 4} U_{\beta 4}|^2 /  (|U_{e4}|^2 + |U_{\mu 4}|^2 + |U_{\tau 4}|^2)$  as a function of $m_4$, along with a direct comparison to existing constraints from the LHC. 
This assumes that $(\nu)$SMEFT operators are present and dilute the constraints on the mixing angles from 
the indirect probes (ii)-(v) above.

\item We subsequently discuss the more restrictive model in which $n=1$ and the $(\nu)$SMEFT operators are ignored,  showing all constraints, including the ones from 
probes (ii)-(v), in the $ |U_{\alpha 4} U_{\beta 4}| -$  $m_4$ plane.
For the latter analysis, we rely in part on recent global analyses~\cite{Atre:2009rg,Drewes:2015iva,Abada:2017jjx,Bolton:2019pcu, Blennow:2023mqx,Fernandez-Martinez:2023phj}, with updates provided in   Appendix~\ref{app:details}, 
which include an improved evaluation of the strongest bound on $|U_{e4}|$, arising from $0 \nu \beta \beta$
decay. 
\end{itemize}

\subsection{$(\nu)$SMEFT  scenario}
\label{sect:results1}

\begin{figure}
\begin{center} 
\includegraphics[width=0.6\textwidth]{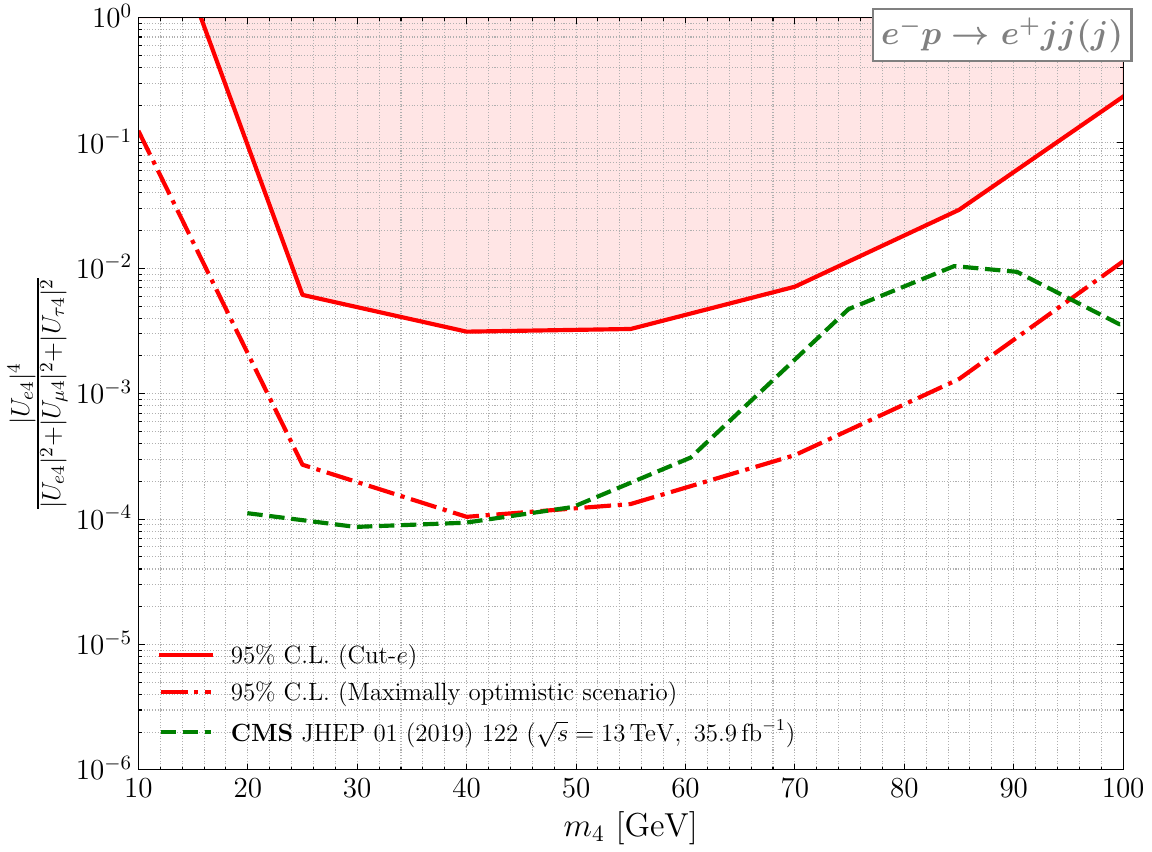}
\vspace{-4ex}
\end{center} 
\caption{Expected 95\% C.L. exclusion limits from the electron channel $e^-p\to e^+jj(j)$ at the EIC with $\sqrt{s}=141\,\mathrm{GeV}$ ($E_e\times E_p=18\,\mathrm{GeV}\times275\,\mathrm{GeV}$), integrated luminosity $\mathcal{L}=100\,\mathrm{fb}^{-1}$, and electron polarization $P_e=70\%$. A charge-misidentification probability $P_{\rm misID}=10^{-3}$ is assumed. The solid red curve shows the limits obtained after applying the kinematic cuts (Cut-$e$) of Box~\ref{box:cut_e}, while the red dot-dashed curve corresponds to a maximally optimistic scenario with complete background rejection and perfect signal reconstruction (see text for details). The dashed green curve corresponds to a CMS analysis at the LHC probing the same coupling combination through resonant production in $\ell$ + jets final states, using a comparable analysis strategy~\cite{CMS:2018jxx}.}
\label{fig:EIC_95CLlimits_e}
\end{figure}

\begin{figure}
\begin{center} 
\includegraphics[width=0.6\textwidth]{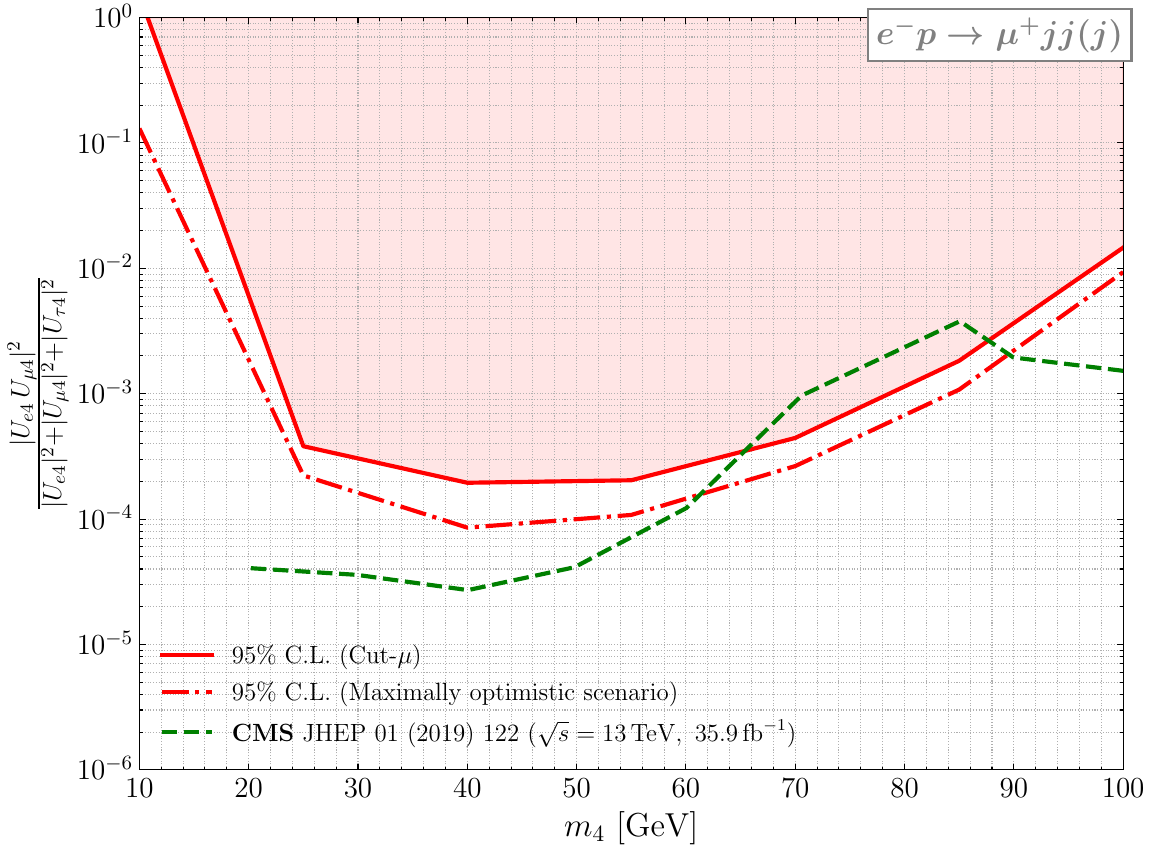}
\vspace{-4ex}
\end{center} 
\caption{Expected 95\% C.L. exclusion limits from the muon channel $e^-p\to \mu^+jj(j)$, assuming an integrated luminosity $\mathcal{L}=100\,\mathrm{fb}^{-1}$, and following the same conventions as in Fig.~\ref{fig:EIC_95CLlimits_e}.
The dashed green curve corresponds to a CMS analysis at the LHC probing the same coupling combination through resonant production in $\ell$ + jets final states, using a comparable analysis strategy~\cite{CMS:2018jxx}.}
\label{fig:EIC_95CLlimits_mu}
\end{figure}

\begin{figure}
\begin{center} 
\includegraphics[width=0.6\textwidth]{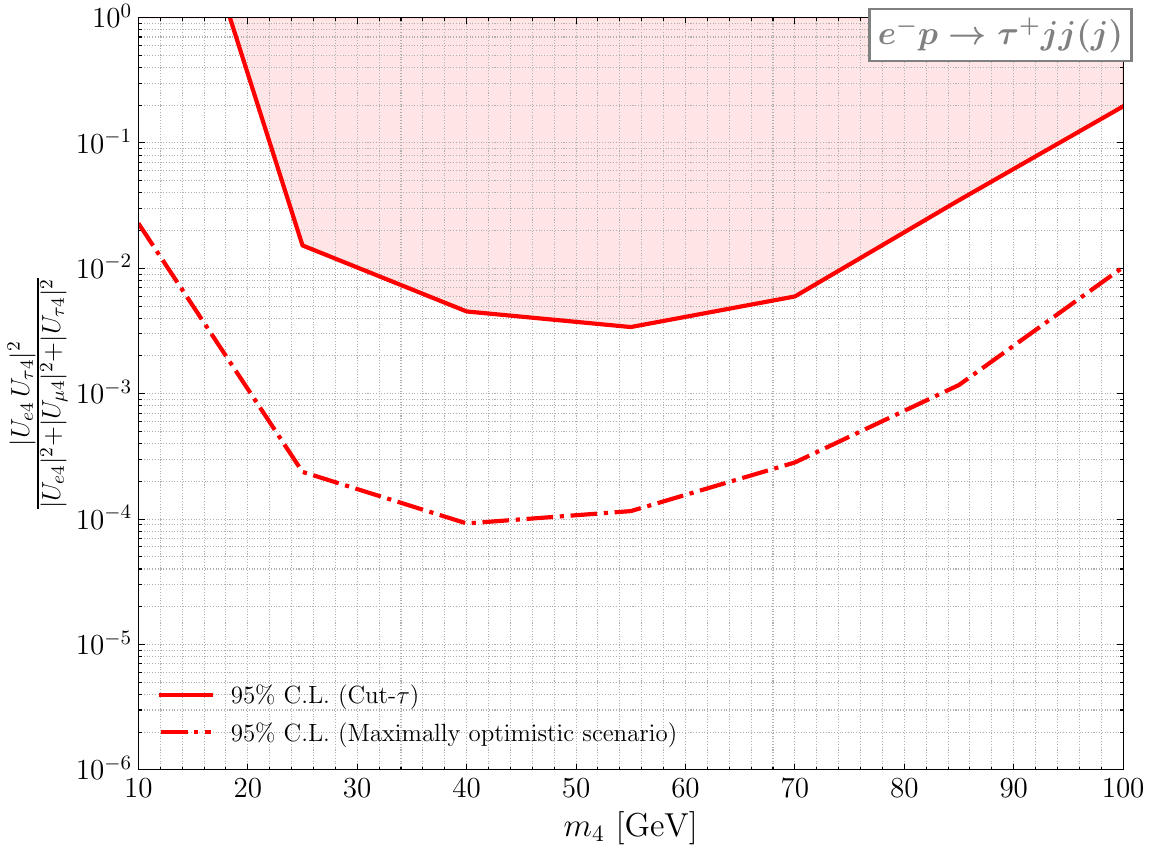}
\vspace{-4ex}
\end{center} 
\caption{Expected 95\% C.L. exclusion limits from the $\tau$ channel $e^-p\to \tau^+jj(j)$, assuming an integrated luminosity $\mathcal{L}=100\,\mathrm{fb}^{-1}$, and following the same conventions as in Fig.~\ref{fig:EIC_95CLlimits_e}.}
\label{fig:EIC_95CLlimits_tau}
\end{figure}

From the kinematic requirements summarized in Boxes~\ref{box:cut_e}--\ref{box:cut_tau}, we obtain the expected signal and background number of events, $N_s$ and $N_b$, respectively, for each lepton-flavor channel. These event counts are then used to derive exclusion limits on the coupling strength $|U_{e4}U_{\alpha 4}|^2/U_4^2$ ($\alpha\in\{e,\mu,\tau\}$) as a function of the HNL mass $m_4$.

Assuming the background prediction is known with negligible uncertainty, we compute the upper bound on the signal contribution using a Bayesian approach following Ref.~\cite{ParticleDataGroup:2020ssz}. For an observed number of events $N_{obs}$ and expected background $N_b$, the $(1-\alpha_\text{\tiny CL})$ credibility upper limit on $N_s$ is obtained from
\begin{equation}
1 - \alpha_\text{\tiny CL} = 1 - \frac{\Gamma(1+N_{obs},\, N_b+N_s)}{\Gamma(1+N_{obs},\, N_b)}\,,
\label{eq:alpha_CL}
\end{equation}
where $\Gamma(a,x)$ is the incomplete gamma function. In the sensitivity projections, we assume the background-only hypothesis, $N_{obs}=N_b$ \cite{Cirigliano:2021img}, thereby extracting the expected upper limit on $N_s$ and translating it into a bound on the coupling strength. The 95\% C.L. exclusion curves for each lepton-flavor channel are presented in Figs.~\ref{fig:EIC_95CLlimits_e}--\ref{fig:EIC_95CLlimits_tau}. The results correspond to an EIC setup with $\sqrt{s}=141\,{\rm GeV}$ ($18\,{\rm GeV}\times275\,{\rm GeV}$), $\mathcal{L}=100\,{\rm fb}^{-1}$, and $P_e=70\%$. A charge-misidentification probability $P_{\rm misID}=10^{-3}$ is assumed.

For each lepton-flavor channel, we present the 95\% C.L. exclusion limits under two benchmark scenarios. The solid red curves correspond to the expected sensitivity obtained after applying the kinematic requirements (Cut-$e$, Cut-$\mu$, and Cut-$\tau$) defined in Boxes~\ref{box:cut_e}--\ref{box:cut_tau}. As indicated by the cut-flow results in Table~\ref{tab:cut-flow}, the residual background contribution remains non-negligible 
for the $e$ case, while, for the assumed luminosity, it is reduced to one (less than one) event  in the $\mu$ ($\tau$) case.
In addition, we consider a `maximally optimistic scenario', defined by the assumptions of complete background rejection and perfect signal reconstruction. The corresponding limits are obtained by setting $N_{\rm obs}=N_b=0$ in Eq.~(\ref{eq:alpha_CL}) and retaining all signal events passing the signal selection criteria (Cut-SS) in Box~\ref{box:signal_selection}. For the $\tau$ channel, this scenario implicitly assumes full $\tau$ reconstruction efficiency, including hadronic decay modes, thereby providing an estimate of the highest sensitivity achievable under ideal detector performance.

As shown in Figs.~\ref{fig:EIC_95CLlimits_e}–\ref{fig:EIC_95CLlimits_tau}, the three lepton-flavor channels exhibit comparable sensitivity across the HNL mass range. The reach of the EIC deteriorates for $m_4 \gtrsim 100\,\mathrm{GeV}$, reflecting the kinematic limitations imposed by the center-of-mass energy. A similar degradation occurs for $m_4 \lesssim 10\,\mathrm{GeV}$, where the generator-level $p_{Tj}$ requirements reduce the efficiency for very light HNLs produced near threshold. Since the signal proceeds through resonant production, the visible jet kinematics become increasingly soft at low masses, leading to a loss of acceptance. In the electron channel $e^-p\to e^+jj(j)$, Fig. \ref{fig:EIC_95CLlimits_e}, the presence of residual background (predominantly NC-DIS) reduces the sensitivity by about one order of magnitude compared to the maximally optimistic scenario. 
A smaller charge-misidentification probability would reduce this 
gap, highlighting the importance of a realistic assessment of $P_{\rm misID}$ for LNV searches at the EIC. Conversely, Fig. \ref{fig:EIC_95CLlimits_mu} indicates that the muon channel $e^-p\to \mu^+ jj(j)$ benefits from a particularly clean experimental signature, underscoring the importance of dedicated muon identification capabilities at the EIC \cite{EIC:2ndDetector2023,EIC:2ndDetector2024,EIC:2ndDetector2025}. 
Note that Figs.~\ref{fig:EIC_95CLlimits_e} and \ref{fig:EIC_95CLlimits_mu} also display the existing CMS constraint from a resonant HNL search in $\ell$ + jets final states at the LHC~\cite{CMS:2018jxx}. This analysis probes the same coupling combination using a comparable resonant production topology and analysis strategy in a similar mass range, allowing for a direct comparison between the projected EIC sensitivity and current LHC bounds. Other resonant searches at the LHC (\emph{e.g.}, Refs.~\cite{ATLAS:2015gtp,ATLAS:2019kpx,CMS:2024xdq,ATLAS:2025lva}) either explore different mass regimes, production mechanisms, or constrain alternative combinations of mixing parameters, highlighting the complementarity of the EIC program. Finally, Fig. \ref{fig:EIC_95CLlimits_tau} presents the projected sensitivity in the $\tau$ channel $e^-p\to \tau^+ jj(j)$. Although the analysis also considers the muon final state, the corresponding limits do not follow a simple rescaling by the relevant branching ratios. This is because the signal kinematics differ significantly between the two channels, as reflected in the distinct distributions shown in Figs.~\ref{fig:distributions_muon_bkg} and \ref{fig:distributions_tau_bkg}. A substantial improvement is observed in the maximally optimistic scenario, where hadronic $\tau$ decay modes are assumed to be fully reconstructed in addition to the leptonic ones. This highlights the strong impact that an efficient and dedicated $\tau$ reconstruction program could have on the sensitivity to LNV signals at the EIC.

\subsection{Restricted $3+1$ scenario}
\label{sect:results2}

\begin{figure}
\begin{center} 
\includegraphics[width=0.6\textwidth]{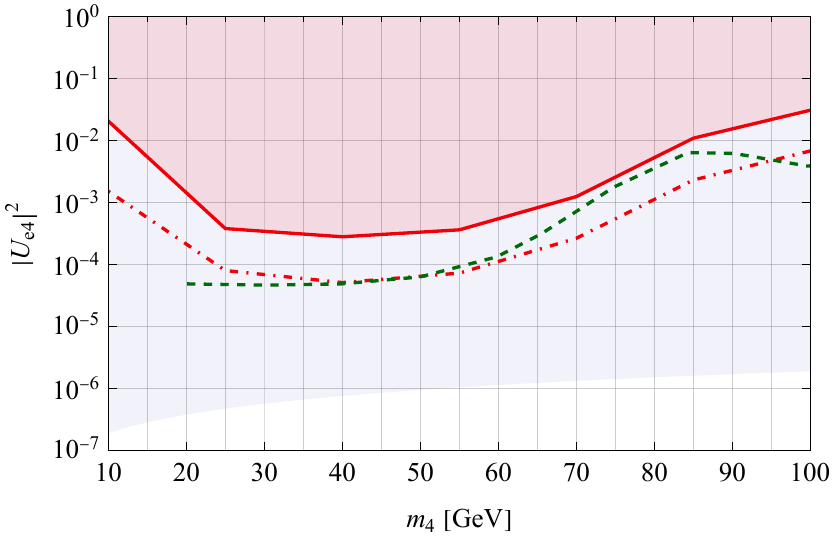}
\vspace{-4ex}
\end{center} 
\caption{The red (green) lines show the expected 95\% C.L. exclusion limits on $|U_{e4}|^2$ from the EIC analysis (the current LHC bound), which have been converted from the bounds depicted in Fig.~\ref{fig:EIC_95CLlimits_e} by
setting $U_4$ to its upper limit,  which provides an  upper bound on the EIC sensitivity.
In the restricted $3+1$ scenario, the shaded blue region is excluded by other 
constraints, dominated in this case by a combination of $0\nu \beta \beta$ decay and tritium $\beta$ decay.
}
\label{fig:ee}
\end{figure}

\begin{figure}
\begin{center} 
\includegraphics[width=0.6\textwidth]{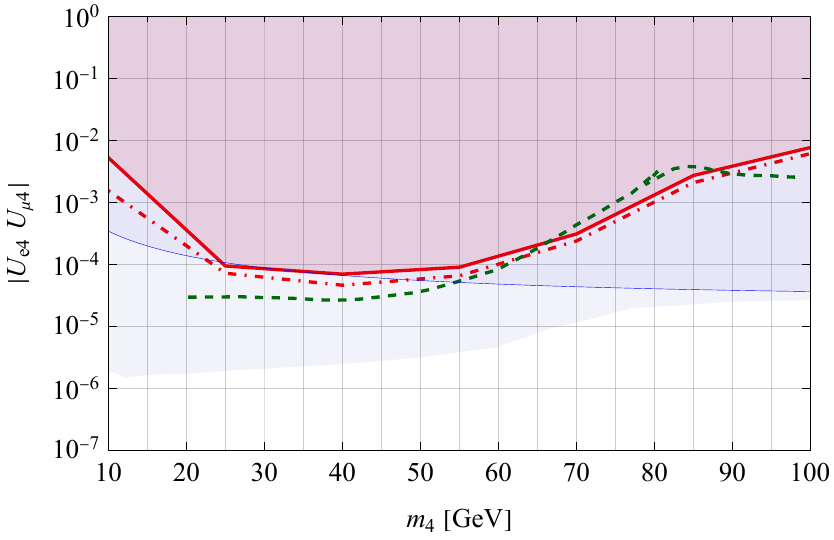}
\vspace{-4ex}
\end{center} 
\caption{The red (green) lines show the expected 95\% C.L. exclusion limits on $|U_{e4}U_{\mu4}|$ from the EIC analysis (the current LHC bound), which have been converted from the bounds depicted in Fig.~\ref{fig:EIC_95CLlimits_mu} by
setting $U_4$ to its upper limit,  which provides an upper bound on the EIC sensitivity.
The shaded blue region is excluded by combining 
constraints on $|U_{e4}|$ (dominated by  a combination of $0\nu \beta \beta$ decay and tritium $\beta$ decay) and $|U_{\mu4}|$ (dominated by LEP and LHC searches), 
while the thin solid line shows the constraint from $\mu\to e$ conversion, which directly probes the combination $|U_{e4}U_{\mu4}|$. }
\label{fig:emu}
\end{figure}

\begin{figure}
\begin{center} 
\includegraphics[width=0.6\textwidth]{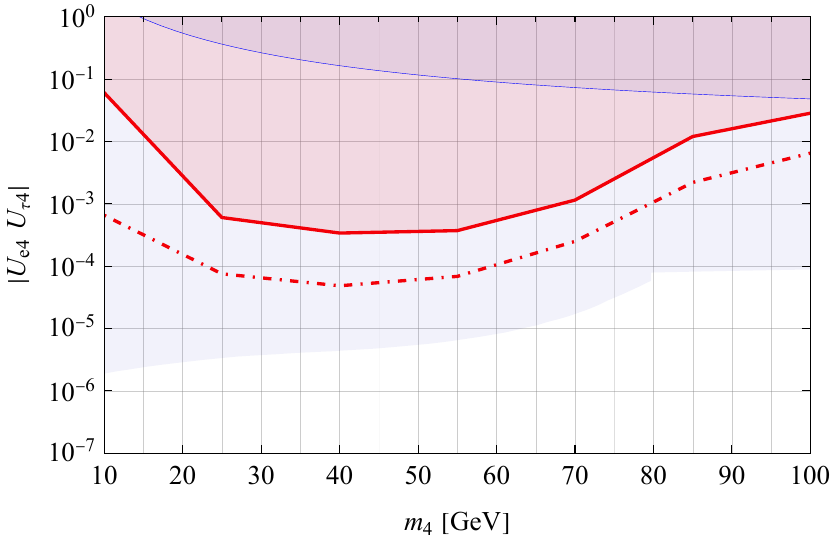}
\vspace{-4ex}
\end{center} 
\caption{The red lines show the expected 95\% C.L. exclusion limits on $|U_{e4}U_{\tau4}|$ from the EIC analysis, which have been converted from the bounds depicted in Fig.~\ref{fig:EIC_95CLlimits_tau} by 
setting $U_4$ to its upper limit,  which provides an upper bound on the EIC sensitivity.
The shaded blue region is excluded by combining 
constraints on $|U_{e4}|$ (dominated by  a combination of $0\nu \beta \beta$ decay and tritium $\beta$ decay) and $|U_{\tau4}|$ (dominated by $Z$ decays), 
while the thin solid line shows the constraint from $\tau\to e\gamma$, which directly probes the combination $|U_{e4}U_{\tau4}|$.}
\label{fig:etau}
\end{figure}

In the restricted case, 
for fixed $m_4$, 
the relevant observables can be parameterized in terms of the following combinations of $U_{e4},  U_{\mu4}, U_{\tau 4}$,
\beq
|U_{e4}|^2, \quad 
|U_{e4} U_{\mu 4}|, \quad
|U_{e4} U_{\tau 4}|, \quad
|U_{\mu 4} U_{\tau 4}|, \quad
U_4 ~. 
\eeq
In order to directly compare EIC (and LHC) constraints to the other probes, we plot (products of) the mixing angles versus $m_4$ for fixed values of $U_4$, 
set to the maximum allowed value. To obtain the latter, we combine the bounds on the individual $U_{\alpha4}$.
For $\alpha=e$, the limit is dominated by  $0\nu\beta\beta$ decay combined with measurements of tritium $\beta$ decays, as discussed in Appendix\ \ref{app:0nubb}. 
For $\alpha=\mu$ the bounds are set by the combination of ATLAS \cite{ATLAS:2025lva} and CMS \cite{CMS:2024xdq} constraints on $|U_{\mu4}|^4/U_4^2$ and the LEP search for $Z$ decays into a HNL and a light neutrino~\cite{DELPHI:1996qcc} for most of the mass range, while weak decays dominate for $m_4\gtrsim m_W$, see Appendices\ \ref{app:weak} and \ref{app:Zdecay}, respectively. Instead, the LHC searches do not play a role in the case of $\alpha=\tau$.
As a result, the combined constraint on $U_4$ is determined mostly by $Z$ decays up to $m_4\simeq m_W$, above which the limit from weak decays takes over.
To show the EIC limits on the combinations $|U_{e4}U_{\alpha4}|$, we multiply the constraints shown in Figs.~\ref{fig:EIC_95CLlimits_e}-\ref{fig:EIC_95CLlimits_tau},  by the bound on $U_4^2$ and take the square root. 
We emphasize that this gives rise to a conservative constraint, since we used the upper limit on $U_4$. 

We plot the resulting 95\% C.L.\ limits in Figs.\ \ref{fig:ee} - \ref{fig:etau}. The EIC bounds are again shown in red, while the limits from 
other probes are depicted by the blue shaded regions. These are obtained by combining the most stringent limits on the separate mixing-matrix elements, $U_{\alpha4}$, similar to the procedure for $U_4$ mentioned above. In addition, the solid thin lines in Figs.\ \ref{fig:emu} and \ref{fig:etau} show the limits from $\mu\to e$ conversion and $\tau\to e\gamma$ that directly probe the flavor-changing combinations, $|U_{e4}U_{\mu4}|$ and $|U_{e4}U_{\tau4}|$, respectively (see Appendix\ \ref{app:LFV} for details).
In all cases, at face value the other probes tend to be more sensitive than the 
projected upper bound from the EIC by an order of magnitude or more. 
On the other hand, the latter have a comparable sensitivity to $\mu\to e$ conversion in the range $m_4=(20- 60)$ GeV and are more sensitive than $\tau\to e\gamma$.

\section{Conclusion and outlook}
\label{sect:conclusion}

The search for  phenomena that violate the conservation of   lepton number  is  well motivated, and a non-zero signal would have implications for the understanding of neutrino mass 
generation and broader BSM physics, including the generation of the cosmic baryon asymmetry. 
In this manuscript, we have explored the prospects of detecting  LNV at the 
EIC through resonant production of a  heavy Majorana  neutral lepton ($N_4$),  
 $e^- p \to N_4 + X$,  followed by $N_4 \to \ell^+_\alpha + W^*$    ($\alpha \in \{e,\mu, \tau \}$),  
 with the $W$ decaying into hadrons, resulting in lepton number and possibly lepton flavor violating 
signatures   $e^- p \to  \ell^+_\alpha + k\, j+X$  where $k=2,3$ denotes the number of jets. 
  
We work in the theoretical framework of the $(\nu)$SMEFT up to dimension seven and 
assume the existence of $n$ HNLs, one of which is in the mass range $10-100$~GeV,  
within the kinematic reach of the EIC.    In this context, HNLs can be produced at the EIC via mixing with light neutrinos 
(see Fig.~\ref{fig:signal_feynman}) or via higher-dimensional operators.  In this exploratory study, we focused 
 on the mixing mechanism, assessed the sensitivity of the EIC, and compared it to other probes of HNLs. 

In Section~\ref{sect:analysis}, we have presented a detailed study of the HNL-induced process 
$e^-p\to \ell_\alpha^++k\,j+X$
and the corresponding Standard Model backgrounds (see Figs.~\ref{fig:distributions_electron_bkg},  \ref{fig:distributions_muon_bkg}, 
and \ref{fig:distributions_tau_bkg}).  
The kinematic distributions presented in Fig.~\ref{fig:distributions_electron_bkg} for the electron flavor channel 
are somewhat different from the ones presented in Ref.~\cite{Batell:2022ogj}. 
This difference originates from our more stringent generator-level cuts on the jet $p_T$ and 
our use of parton shower and detector response, which significantly extends the parton-level analysis of Ref.~\cite{Batell:2022ogj}. 
Based on the signal and background distributions, we have suggested a set of cuts  
for signal selection and background minimization (see Boxes~\ref{box:signal_selection}--\ref{box:cut_tau}), which provide the basis for a first sensitivity study.

Our analysis for the $\mu$ and $\tau$ channels $e^-p\to \ell_\alpha^++k\,j+X$ ($\alpha\in\{\mu,\tau\}$)
{\it assumes} a muon detection efficiency comparable to the electron detection efficiencies. 
We emphasize here that the muon channel $e^-p\to \mu^+ jj(j)$ 
benefits from a particularly clean experimental signature. This underscores the importance of muon identification  
 capabilities at the EIC \cite{EIC:2ndDetector2023,EIC:2ndDetector2024,EIC:2ndDetector2025}, 
 to extend the  reach of this machine to BSM physics. 
Our analysis of the  $\tau$ channel at the moment relies on reconstructing the $\tau$ through its 
leptonic decay into a muon and neutrinos. 
The   muon $p_T$ distribution  resulting from the 
$e^-p\to \tau^+ jj(j)$  signal events is softer than the one from 
$e^-p\to \mu^+ jj(j)$,  and thus the $\tau$ channel 
suffers from intrinsic Standard Model backgrounds.  Reconstruction of the $\tau$ from hadronic decays can 
dramatically improve the sensitivity in the $\tau$ channel, through background reduction and larger branching fractions. 

In Section~\ref{sect:pheno} we illustrated the  reach of the EIC in probing HNL mixing angles in the mass range $m_4 \in [10,100]$~GeV. 
We first considered a scenario in which indirect constraints from low-energy probes can be canceled by $(\nu)$SMEFT operators 
(Figs.~\ref{fig:EIC_95CLlimits_e} - \ref{fig:EIC_95CLlimits_tau}). 
Our results show that the EIC with muon detection capabilities and an integrated luminosity of  $100~\mathrm{fb}^{-1}$ 
can be competitive with the LHC~\footnote{We base this conclusion on the existing analysis of Ref.~\cite{CMS:2018jxx} and projecting improvements that scale with the inverse square root of the integrated luminosity.}.
We then considered a more restrictive $3+1$ scenario with no $(\nu)$SMEFT contributions (Figs.\ \ref{fig:ee} - \ref{fig:etau}).  
In this case constraints from other probes are stronger than the projected upper 
bound on the sensitivity from the EIC by an order of magnitude or more. 
On the other hand, the EIC  processes  have a comparable sensitivity to $\mu\to e$ conversion 
in the range $m_4=(20- 60)$ GeV and are more sensitive than $\tau\to e\gamma$.

Our exploratory study  motivates further experimental   
 assessment of  both  muon detection capabilities  and $\tau$ hadronic reconstruction, 
 as well as further theoretical investigations.  
In particular, our findings motivate a more general theoretical analysis in which the  $(\nu)$SMEFT operators that can interfere 
with resonant production at colliders (LHC, EIC) and with the strongest probes at lower energies 
are simultaneously turned on. In this more general $(\nu)$SMEFT framework, the EIC can play a central role. 
In fact, for resonant production at colliders, given the different energy scales involved, we expect that the relative importance of the mixing mechanism and dim-6 and dim-7 $(\nu)$SMEFT operators will differ between the EIC and the LHC. 
Therefore, given that we find constraints from the EIC and  the  LHC 
to be  within $\mathcal{O}(1)$,   
searches in all these channels 
will provide multiple paths to discovery and 
ways to disentangle the origin of a possible positive signal or fully constrain the parameter space in case of null results.

\acknowledgments 
We thank Zuhal Demiroglu, Abhay Deshphande,  Ciprian Gal, Charlotte Van Hulse, Krishna Kumar, Ming Liu, Rafael Coelho Lopes de Sá, Frank Petriello, Richard Ruiz, and Zhoudunming Tu for valuable 
discussions and correspondence on various aspects of this work. 
V.C., W.D., and S.U.Q. gratefully acknowledge support from the  U.S.\ DOE under Grant No.\ DE-FG02-00ER41132. 
E.M. gratefully acknowledges financial support by Los Alamos
National Laboratory's Laboratory Directed Research and Development program under project
20250164ER.  Los Alamos National Laboratory is operated by Triad National Security, LLC,
for the National Nuclear Security Administration of U.S.\ Department of Energy (Contract No.\
89233218CNA000001). K. F. acknowledges the support by RIKEN iTHEMS SUURI-COOL (KEK) program.

\appendix 

\section{Constraints on HNLs from other processes}
\label{app:details}

\subsection{LEP and LHC constraints}\label{app:Zdecay}
The decay of the $Z$ boson to a HNL provides strong  constraints on 
$
\sum_{i<4}\Big|\sum_\alpha U_{\alpha i}^*  U_{\alpha 4} \Big|^2
\simeq \sum_\alpha
|U_{\alpha 4}|^2$, where $\alpha = \{e, \mu, \tau \}$. 
The non-observation of $Z \to \nu_i N_4$ with $i<4$ at LEP~\cite{DELPHI:1996qcc}
bounds  this combination  at the level of $2 \times 10^{-5}$  for $m _4 < 80$~GeV. 
In addition,  CMS and ATLAS have performed searches for $W$ production followed by $W\to N_4\ell$ \cite{CMS:2024xdq,ATLAS:2025lva}, which probe the combinations of couplings $|U_{\alpha 4}|^4/U_4^2$. For $\alpha=\mu$, the results, combined with the LEP limits, provide a slightly more stringent bound on $|U_{\mu4}|$.

\subsection{Weak charged-current  decays}\label{app:weak}
Weak decays of  mesons and leptons with masses smaller than $m_4$ constrain the absolute value of mixing angles $|U_{\alpha 4}|$ 
independently of $m_4$. 
The effect of the mixing angles shows up both directly, by contributing to the relevant rates, and indirectly, by modifying the measured Fermi constant. In particular, the Fermi constant extracted from muon decay $G_\mu$ is related to  $G_F$ appearing in semileptonic decays via \cite{Fernandez-Martinez:2016lgt,Blennow:2016jkn} 
\beq
G_\mu^2 = G_F^2 \, (1 - |U_{e4}|^2) (1 - |U_{\mu 4}|^2)~.
\eeq
The combination of direct and indirect effects in the rate of semileptonic decays involving the underlying quarks  $d_j$  $u_i$ and leptons  $\ell_\alpha$,  $\bar \nu_\alpha$ then scales as 
\beq
\Gamma_{u_i, d_j, \ell_\alpha, \nu_\alpha} \, \propto \,  G_F^2 \, |V_{u_i d _j}|^2 \, (1 - |U_{\alpha 4}|^2) 
 \, =  \, G_\mu^2  \,  |V_{u_i d _j}|^2 \,\frac{1 - |U_{\alpha 4}|^2}{(1 - |U_{e4}|^2) (1 - |U_{\mu 4}|^2)}~, 
\eeq
where $V_{u_i d_j}$ denotes a generic element of the Cabibbo-Kobayashi-Maskawa (CKM) matrix. 

The strongest constraints on the mixing angle come from tests of unitarity of the CKM matrix and from lepton universality tests.
For CKM unitarity, the effective channel-dependent CKM elements extracted from the decay  rate 
$\Gamma_{u_i, d_j, \ell_\alpha, \nu_\alpha}$ 
read, for small mixing,  
\beq
\bar V_{u_i d_j} = V_{u_i d_j} \times \left(1 + \frac{1}{2} \left( |U_{e4}|^2 + |U_{\mu 4}|^2 - |U_{\alpha 4}|^2 \right) \right)~.
\eeq
To set constraints, we use extraction of $\bar V_{us}$
from $K \rightarrow \pi \mu \nu$
and $K \rightarrow \pi e \nu$ decays, which are given for electron and muon channels in Refs. \cite{ParticleDataGroup:2024cfk,MoulsonTalk}. 
For the kaon form factor at zero momentum,
$f_+(0)$, we use the average of $N_f = 2 + 1 + 1$ lattice QCD calculations \cite{FlavourLatticeAveragingGroupFLAG:2024oxs,Carrasco:2016kpy,FermilabLattice:2018zqv}.
In addition, we use the ratio $\Gamma(K\rightarrow \mu \nu)/\Gamma(\pi\rightarrow \mu \nu)$, to obtain \cite{FlavourLatticeAveragingGroupFLAG:2024oxs}
\begin{equation}
    \frac{|V_{us}|}{|V_{ud}|} = 0.23126(51)~,
\end{equation}
using the lattice QCD determination of the ratio of $f_K/f_\pi$ \cite{FlavourLatticeAveragingGroupFLAG:2024oxs,Dowdall:2013rya,Carrasco:2014poa,Bazavov:2017lyh,Miller:2020xhy,ExtendedTwistedMass:2021qui}.
Finally, we determine $\bar V_{ud}$ from superallowed $\beta$ decays \cite{Hardy:2020qwl,ParticleDataGroup:2024cfk}
\begin{equation}
    |\bar V_{ud}| = 0.97367(32)~.
\end{equation}
This determination only involves $\alpha = e$, and thus is sensitive to $|U_{\mu 4}|^2$.

Lepton flavor universality tests involve the ratios  
${\cal R}_{\alpha \beta} =  \Gamma_{u_i, d_j, \ell_\alpha, \nu_\alpha}/\Gamma_{u_i, d_j, \ell_\beta, \nu_\beta}$, which can be re-expressed as
\beq
{\cal R}_{\alpha \beta} = {\cal R}_{\alpha \beta} \Big |_{\text{SM}} 
\left(1 - |U_{\alpha 4}|^2 + |U_{\beta 4}|^2 \right) ~.  
\eeq
Similarly, for leptonic 
$\tau$ decays, one finds \cite{Bryman:2021teu}
\begin{eqnarray}
    R^\tau_{\mu/e} &=&\frac{\Gamma(\tau \rightarrow \mu \bar\nu_\mu \nu_\tau )}{\Gamma(\tau \rightarrow e \bar\nu_e \nu_\tau )} =  \left. R^\tau_{\mu/e} \right|_{\text{SM}} \frac{1- |U_{\mu 4}|^2}{1- |U_{e 4}|^2},   \\
    R^\tau_{\tau/\mu} &=& \frac{\Gamma(\tau \rightarrow e \bar \nu_e  \nu_\tau )}{\Gamma(\mu \rightarrow e \bar \nu_e  \nu_\mu )} = \left. R^\tau_{\tau/\mu} \right|_{\text{SM}} \frac{1- |U_{\tau 4}|^2}{1- |U_{\mu 4}|^2}. 
\end{eqnarray}
In our analysis, we use ratios of $\pi$
and $K$ decays to muons and electrons, $D_s$ meson decays to $\tau$ and muons, and leptonic $\tau$ decays. Theoretical input is taken from Refs. \cite{Bryman:2021teu,Pich:2013lsa}.

We perform a fit to lepton universality and CKM unitarity observables, with four fit parameters: 
$\lambda = V_{us}$ and the three mixing angles $U_{\alpha 4}$. We find that the $\chi^2$ of the fit does not improve under the SM assumption, implying that adding HNLs  to the theory does not resolve the Cabibbo anomaly. 
At the 95\% confidence level, we obtain 
\begin{align}
|U_{e4}|^2 &<3.7 \times 10^{-3} \\
|U_{\mu4}|^2 &<3.8 \times 10^{-4} \\
|U_{\tau4}|^2 &<4.1 \times 10^{-3}.
\end{align}

\subsection{LFV and LNV  decays}\label{app:LFV}

The rate for the decays  (LFV or LNV) of  mesons,   leptons, or the $Z$ boson,
  involving flavors $\alpha$ and $\beta$   takes the form 
\beq
\Gamma^{(i)}_{\alpha \beta}  \, =  \,  \gamma^{(i)}    \, (|U_{\alpha 4} | | U_{\beta 4}|)^2   \, g^{(i)} (m_4)~. 
\eeq
The LNV decays are induced by dimension-9 operators that arise at tree level, leading to the scaling $g^{(i)} = 1/m_4^2$,  while $g^{(i)}$ is a function of  $(M_W/m_4)$ for LFV decays, which arise a loop-level.
Below, we discuss the resulting constraints. 

\begin{itemize}

\item {\bf LFV $\mu$ and $\tau$ decays:} 
In the $\mu-e$ sector, the strongest constraints arise from 
$\mu \to e \gamma$ and $\mu \to e$ conversion in nuclei.  
Using results from Ref.~\cite{Alonso:2012ji}, where all the relevant decay rates have been computed, 
we update the limits. We find that, for the mass range of interest, the strongest constraints arise from $\mu \to e$ conversion in 
gold~\cite{SINDRUMII:2006dvw} and titanium~\cite{SINDRUMII:1993gxf}, followed by $\mu \to e \gamma$~\cite{MEGII:2023ltw}.
In the $\tau-e$ sector, many LFV decay modes are available. 
We take as a representative the radiative mode $\tau \to e \gamma$ for which a 90\% CL upper limit 
of $3.3 \times 10^{-8}$ has been set in Ref.~\cite{BaBar:2009hkt}.

\item {\bf $\mu^- \to e^+$ conversion:}  Here the theoretical uncertainties are larger. Ref.~\cite{Berryman:2016slh} provided an estimate of the decay rate based on dimensional analysis, in terms of an undetermined scale $Q \sim m_\mu$.
Using this approach, we obtain  the weak bound 
\beq
|U_{e4} U_{\mu 4}| \leq 4.25  \ \left( \frac{m_\mu}{Q} \right)^3  \frac{m_4}{{\rm GeV}}~.
\eeq
On the other hand, Ref.~\cite{Domin:2004tk} performed  a calculation of the nuclear matrix elements induced by  the exchange of 
Majorana neutrinos (light or heavy).  Using their results, we obtain the even  weaker limit 
\beq
|U_{e4} U_{\mu 4}| \leq 3 \times 10^{4} \, \frac{m_4}{{\rm GeV}}~.
\eeq

\item {\bf LNV  $\tau$  decays:}  
Ref.~\cite{Liao:2021qfj} studied LNV $\tau$ decays induced by dimension-7 in the SMEFT, which in turn generate  dim-6 and dim-9 operators in the LEFT. 
Using the results of Ref.~\cite{Liao:2021qfj} for the hadronic matrix element,  from the process $\tau^\pm  \to e^\mp  \pi^\pm \pi^\pm$  we obtain the very loose bound 
\beq
|U_{e4} U_{\tau 4}| \leq 6.2 \times 10^3 \, \frac{m_4}{{\rm GeV}}~.
\eeq

\end{itemize}

\subsection{Neutrino-less double beta decay}\label{app:0nubb}

We compute the decay rate due to LNV from both light neutrinos and $N_4$, which induces a dimension-9 operator at low energy~\cite {Cirigliano:2018yza}.  Treating the $N_4$ and $m_{\beta \beta}$  contributions as independent, 
and focusing on the $^{136}$Xe isotope for which the strongest half-life bound is currently available~\cite{KamLAND-Zen:2024eml},
we obtain~\cite{Cirigliano:2021peb} 
\begin{equation}
\left[ T^{0\nu}_{1/2}\left(^{136}{\rm Xe}\right)\right]^{-1}  = 
\Bigg[
a \, |\bar m_{\beta \beta}|^2 + 
b \,  \frac{|m_{\beta \beta}| |U_{e4}|^2}{\bar m_4} \cos \eta 
 +  
 c \, \left( \frac{|U_{e4}|^2}{\bar m_4}  \right)^2
 \Bigg] \, {\rm yr}^{-1}~,
 \label{eq:nldbd}
\end{equation}
where $\bar m_{\beta \beta} = m_{\beta \beta}/{\rm eV}$, 
$\bar m_{4} = m_{4}/{\rm GeV}$, 
and $\eta$ is the unconstrained relative phase between $m_{\beta \beta}$ and $U_{e4}^2$.
Using the nuclear matrix elements of Ref.~\cite{Hyvarinen:2015bda} and the phase space factors extracted from Ref.~\cite{Horoi:2017gmj} 
one obtains $a= 2.4 \times 10^{-24} $, $b= 1.5 \times 10^{-16}  $, $c=2.3 \times 10^{-9}$.
The strongest constraint on the $0\nu\beta\beta$ decay half-life comes from the 2024 KamLAND-Zen measurement \cite{KamLAND-Zen:2024eml}, which implies the lower limit $T^{0\nu}_{1/2}({}^{136} \mathrm{Xe})>4.4\times 10^{26}$ yr at 95\% C.L.

\begin{figure}
\begin{center} 
\includegraphics[width=0.48\textwidth]{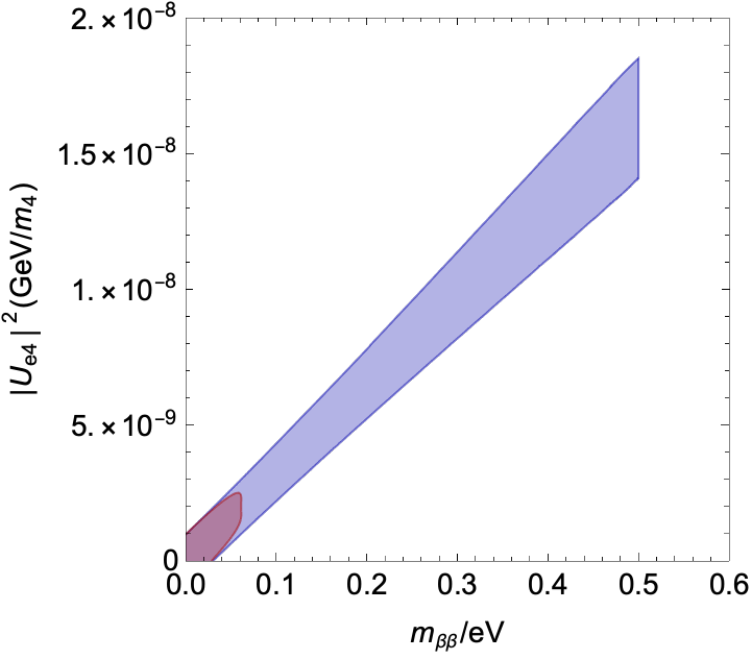}
\end{center} 
\caption{Allowed region obtained from the combination of $0\nu\beta\beta$ constraints with the KATRIN (cosmological) bound in blue (red). After marginalizing over the phases, the limits from KATRIN and cosmology provide an upper limit on $m_{\beta\beta}$. The contours are obtained from $\chi^2 = \left(T^{0\nu}_{1/2}/T^{0\nu\, ({\rm expt})}_{1/2}\right)^2+ \left(m_{\beta\beta}/m_{\beta\beta}^{(\rm expt)}\right)^2$, where $m_{\beta\beta}^{(\rm expt)}$ refers to the upper limits from either the KATRIN experiment or cosmology.}
\label{fig:nldbd2}
\end{figure}

For $ \cos \eta >0$, marginalizing over $m_{\beta \beta}$ we obtain $|U_{e4}|^2 < 1.1 \times 10^{-9} \, m_4/{\rm GeV}$, where the upper limit is reached 
for $m_{\beta \beta} \to  0$. 
For $\cos \eta <0$  cancellations are possible. 
In particular, for $\cos^2 \eta < \cos^2 \eta_0 = 4 a c/b^2$ ($=0.98133$ with the matrix elements we use)  
a flat direction opens up, implying the contributions to $0\nu\beta\beta$ are canceled when
$|U_{e4}|^2/\bar m_4  <  b/(2 c) (1 + \sqrt{1 - 4  c a /b^2}) \ |\bar m_{\beta \beta}| \approx 3.8 \times 10^{-8} \ |\bar m_{\beta \beta}|$.
While  this could in principle relax the bound on $|U_{e4}|^2/m_4$, in practice 
the effect is minimal,  given the small coefficient $\sim 10^{-8}$ and the fact that  
$|m_{\beta \beta}|$ itself is indirectly bound by tritium beta decay and cosmology, 
through its correlation with $m_\beta = \sqrt{\sum_{i=1}^3 |U_{ei}|^2 m_i^2}$ and $\Sigma = \sum_{i=1}^3 m_i$. 
The current bounds on $m_\beta$~\cite{KATRIN:2024cdt} and $\Sigma$~\cite{DESI:2024mwx} roughly translate into upper bounds $|m_{\beta \beta} | < 0.5$~eV 
and $|m_{\beta \beta} |< 0.061$~eV, respectively~\footnote{For $m_4$ in the 10-100 GeV region, 
as considered here, the contribution of the Heavy Neutral Lepton does not impact 
the analysis of tritium decay and the cosmological signatures of light neutrino mass.}. 
We show the allowed region in the parameter space of  $|U_{e4}|^2$ vs $m_4$ in Fig.\ \ref{fig:nldbd2}. Here, the blue (red) shaded region denotes the combination of $0\nu\beta\beta$ and the KATRIN measurement (cosmological constraints). 
The bound on the HNL coupling and mass relaxes by at most one order of 
magnitude compared to the $|m_{\beta \beta}|=0$ case. 
Using only the KATRIN constraint on $|m_{\beta \beta}|$ leads to 
$|U_{e4}|^2 < 1.5 \times 10^{-8} \, m_4/{\rm GeV}$. 
We used the combination of $0\nu\beta\beta$ and the constraints from KATRIN to obtain the sensitivity plots in the main text.

\bibliographystyle{JHEP} 
\bibliography{bibliography}

\end{document}